\documentclass{aastex}
\usepackage{rotating}
\usepackage{amssymb}
\def \lp{\>\> .}
\def \lc{\>\> ,}

\def \c2{cm$^{-2}$}

\def \nh3{NH$_3$}
\def \n2h{N$_2$H$^+$}

\def \nh2{n_{H_2}}
\def \nh1{n_{HI}}

\def \th{$^{13}$CO}

\def \h2{H$_2$}

\def \H2{H$_{2}$}

\def \be{\begin{equation}}
\def \ee{\end{equation}}
\def \bf{\begin{figure}}
\def \ef{\end{figure}}

\shorttitle{Radial Power Laws in Molecular Cloud Cores}
\shortauthors{Kr\v{c}o et al.}

\begin{document}

\title{Geometry-Independent Determination of Radial Density Distributions in Molecular Cloud Cores and Other Astronomical Objects}

\author{Marko Kr\v{c}o}
\affil{Department of Astronomy, Cornell University, Ithaca, NY 14853}
\email{marko@astro.cornell.edu}

\and

\author{Paul F. Goldsmith}
\affil{Jet Propulsion Laboratory, Pasadena, CA 91109 and Cornell University, Ithaca NY 14853}

\begin{abstract}
We present a geometry-independent method for determining the shapes of radial volume density profiles of astronomical objects whose geometries are unknown, based on a single column density map.
Such profiles are often critical to understand the physics and chemistry of molecular cloud cores, in which star formation takes place. 
The method presented here does not assume any geometry for the object being studied, thus removing a significant source of bias.
Instead it exploits contour self-similarity in column density maps which appears to be common in data for astronomical objects.
Our method may be applied to many types of astronomical objects and observable quantities so long as they satisfy a limited set of conditions which we describe in detail. 
We derive the method analytically, test it numerically, and illustrate its utility using 2MASS-derived dust extinction in molecular cloud cores. 
While not having made an extensive comparison of different density profiles, we find that the overall radial density distribution within molecular cloud cores is adequately described by an attenuated power law.
\end{abstract}

\section{Introduction}
\label{cintro}

A long--standing problem in astrophysics is a result of our inability to determine the three-dimensional structure of distant objects.
This limitation has often inhibited our understanding of the internal structure of even relatively well--defined and isolated astronomical objects, such as molecular cloud cores. 
Assuming that such an object is spherically symmetric, or has some other simple geometry, often permits us to describe the object's internal structure using one or more radial profile functions. 
Such radial profiles are frequently used to examine or model the physics and chemistry which govern such objects.

It is relatively safe to assume that stars and planets are spherical or that a spiral galaxy has a disk and a bulge. 
These assumptions become problematic when studying objects without obvious symmetry. 
Molecular cloud cores exhibit a wide variety of shapes that very rarely resemble any simple geometry. 
Hence, determining their internal structure while using a geometric assumption will always yield some bias in any derived radial profile. 
In this paper we describe a technique which may be used to obtain limited, yet useful information about an object's radial profile function without making any assumptions about the object's shape, orientation, or the nature of the radial profile function. This is done using a single two-dimensional column density map as the entire available data on the source.

A variety of such techniques have been used to determine the radial density distribution in molecular cloud cores (also referred to as dense cores) in studies over more than three decades.  Early work employed \th\ emission \citep{Dickman1983, Arquilla1985}.  Optical extinction was utilized by \citet{Cernicharo1985} to determine the density distribution within a number of dark clouds in Taurus. These techniques were recognized to have weaknesses (inability to trace high column density regions for optical extinction, and variable abundance due to e.g. freezeout for carbon monoxide).  Subsequent efforts have largely moved to measument of stellar reddening in the near-infrared, allowing accurate probing of the extinction to much greater columns \citep{Alves2001a, Alves2001b}.  \citet{Kandori2005} employed measurement of infrared colors and stellar densities to obtain the density structure of 10 dense cores.  The infrared color excess technique was utlized by \citet{Pineda2010} to derive the density distribution in cloud cores in Taurus using 2MASS data. Continuum emission may also be used to study the temperature, and density distribution of dust in the ISM. With Herschel data, \citet{Launhardt13} were able to probe the dust within 12 molecular cloud cores. \citet{Stutz10} used both dust extinction as well as emission to model the column density and temperature distribution of CB244. To introduce this new technique, we will constrain ourselves to dust extinction as it is simpler, and temperature-independent.

In some of the previous work, a singe power law radial density profile was fitted to the data \citep{Arquilla1985, Cernicharo1985}, with exponents typically between 1 and 2 found.  Other studies  used a Bonner--Ebert Sphere  \citep{Bonnor1956, Ebert1955} to model the density profile, which characteristically has a flat density profile in the central region transitioning to a $r^{-2}$ radial dependence towards the edge of the core \citet{Dickman1983, Alves2001a, Alves2001b, Kandori2005}.  A function with a similar form gave a good fit to the data of \citet{Pineda2010}.  

Our technique improves on previous methods by eliminating any geometric assumptions, as well as any a priori assumptions about the nature of the radial profile function.
There are certain limitations to the technique as well as criteria which must be fulfilled.
These are discussed in detail in Section \ref{cassumptions}. 
The most important limitation and constraints may be summarized as follows.
\begin{itemize}
\item Since a two dimensional projection cannot uniquely define a three-dimensional object without additional information, it is impossible to obtain absolute values for the radial profile function without additional information or assumptions. It is however possible to obtain the form of the function which differs from the original profile by two unknown, geometry-dependent scalars.
\item The internal structure of the object in question must be describable using a radial profile function.
\end{itemize}

In theory the technique may be used to study spectral line emission, absorption, continuum emission, extinction, etc. 
It may be applied to any object provided it is consistent with the assumptions described in Section \ref{cassumptions}. We will demonstrate that the technique is useful even in cases where only a portion of an object exhibits contour self-similarity.

To illustrate and validate the technique we have chosen to apply it to maps of the dust extinction in molecular cloud core column density maps derived from 2MASS data on stellar reddening.
This paper is formulated so as to introduce a novel methodology by presenting an analytical derivation, testing it against simulated data, and finally applying it to real data.
Section \ref{cassumptions} describes the initial assumptions which must be fulfilled in order for the technique to be applicable to a given object. 
The assumptions yield critical relationships which illustrate key aspects of this technique. 
Section \ref{cderivation} derives the technique analytically using two different methods. 
Section \ref{cnumeric} applies the technique to a set of simulated data designed to test its validity as well as to expose its performance under a variety of circumstances. 
Section \ref{creal} discusses the use of 2MASS dust extinction maps and applies the technique to several clouds.
We make a comparison with previous methods for measuring radial profiles in Section \ref{ccomparison}.
We discuss the results and the performance of this new technique in Section \ref{cdiscussion}.

\section{Assumptions}
\label{cassumptions}

The goal of this research is to extract the maximum available information regarding the internal volume density structure of an object using a single column density map observed from one line of sight direction, while making the fewest possible assumptions. 
We show how it is possible under certain conditions to obtain the form of an object's volume density profile function without assuming a specific geometry, or making any assumptions about the function that governs the radial density profile. 
To this end it is necessary to detail the assumptions used in this work.

The method described here only relies on the three assumptions below which are made for all cases.

\begin{description}
\item[Assumption 1:] The object studied must be optically thin in whatever observable quantity is being measured in the sense that
\begin{equation}
\label{csimplen}
	N(y,z) = \int_{-\infty}^{\infty} n(x,y,z) dx \lc 
\end{equation}
	where $N(y,z)$ is the measured column density at position $(y,z)$ and $n(x,y,z)$ represents the volume density at position $(x,y,z)$. 
Throughout this paper, the x axis is arbitrarily chosen to represent the line of sight direction.

\item[Assumption 2:] The volume density of the object can be entirely characterized using a single function that describes the volume density profile.
\end{description}
	
        The following can be considered to follow from assumption 2.

\begin{description}
\item[Assumption 2a:] Any object which satisfies assumption 2 must be described using two functions; 
One describes the cloud's geometry, while the second describes its radial volume density profile. 
\end{description}

We define the object's shape using a core function
	\begin{equation}
		a(\alpha, \theta)=a_c f_c(\alpha,\theta) \lc
	\end{equation}
	where $a(\alpha,\theta)$ has units of length and describes the size of the object's core along each direction originating from the object's center. $a_c$ is a constant with units of length, and $f_c(\alpha,\theta)$ is a dimensionless function which scales the core radius along each $(\alpha,\theta)$ to produce a shape for the object. 
In the case of a sphere, $f_c(\alpha,\theta)=1$ while $a_c$ represents the radius.
Spherical coordinates are chosen here to emphasize the fact that the core function depends only on direction from the object's center, and not on distance.
	
When working with arbitrary shapes it is convenient to define a new, dimensionless parameter $r_{rc}$ which is equal to the ratio between the distance from some point $(x,y,z)$ to the object's center, and the core radius $a(\alpha,\theta)$ along the same direction. 
With the object's center located at the origin of the coordinate system $(0,0,0)$,
	\begin{equation}
	\label{corirrc}
		r_{rc}(x,y,z)= \frac{\sqrt{x^2 + y^2 + z^2}}{a(\alpha,\theta)} \lp
	\end{equation}
	 For a sphere, $r_{rc}(x,y,z)=\sqrt{x^2 + y^2 + z^2}/a_c$. 
We commonly refer to the surface described by $r_{rc}=1$ as the core.
In order to fully describe the geometry of an object which meets assumption 2, $f_c(\alpha,\theta)$ must describe a closed surface such that a vector from the object's center along any direction will cross the surface exactly once. 
This permits the definition of a radial volume density function that is dependent on $r_{rc}$ and governs the volume density distribution of the entire object. We define
	\begin{equation}
	\label{corin}
	n(x,y,z) = n_0 f_n(r_{rc}) \lc
	\end{equation}
	where $n(x,y,z)$ represents the volume density at position $(x,y,z)$, and $f_n(r_{rc})$ is a dimensionless function that governs the radial volume density profile. 
$n_0$ is a constant representing the volume density where $f_n(r_{rc})=1$. 
	 	 
$r_{rc}(x,y,z)$ and $n(x,y,z)$ can fully characterize any object which satisfies assumption 2. 
Neither $r_{rc}(x,y,z)$, nor $n(x,y,z)$ can ever be fully determined from a single column density map using only one observable quantity without additional information, since a column density map in and of itself can not uniquely define a three-dimensional object.
It is possible to determine the function $f_n$ as well as certain properties of $a(\alpha,\theta)$ by taking advantage of the self-similarity imposed on the object by assumption 2. 
Any object which satisfies assumption 2 satisfies the implied assumptions below.

\begin{figure}
\includegraphics[scale=.75]{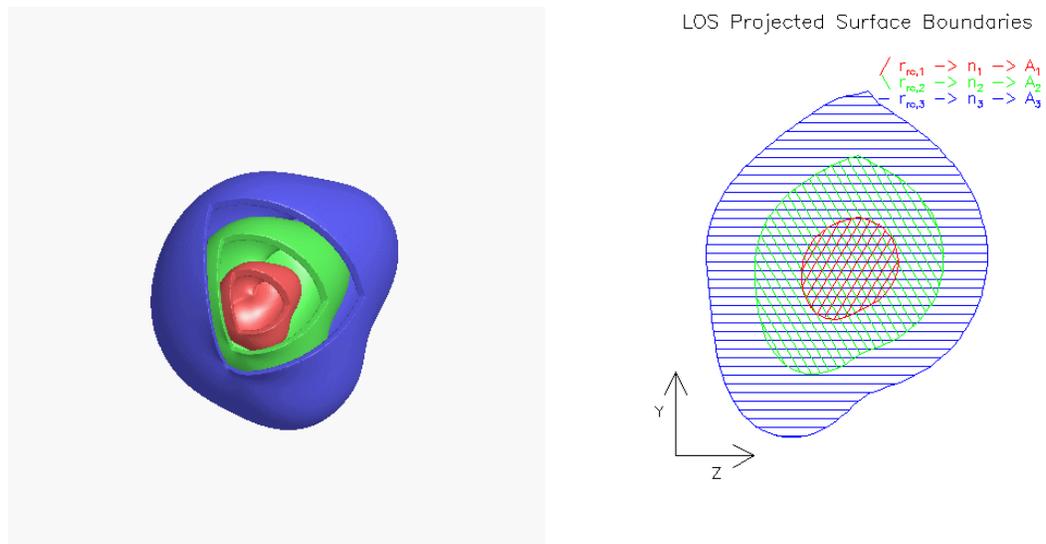}
\caption{(Left) A three-dimensional representation of surfaces with three specific values of $r_{rc}$ for an arbitrary object which satisfies assumption 2. Each surface shares the same shape and orientation, while differing only in scale. (Right) Projections of the three surfaces along the line of sight (LOS). Each surface is characterized by a specific value of $r_{rc}$, its projected Area ($A$), and its volume density ($n$) as in Equation \ref{corin}. Projected areas have the same shape and orientation, while differing only in scale. Since the object is assumed to be optically thin, the projected column densities from each surface add linearly to produce the total column density. 
\label{cfig1}}
\end{figure}

\begin{description}
\item[Assumption 2b:] Specific values of $r_{rc}$ describe three-dimensional surfaces of equal volume density. The left panel of Figure \ref{cfig1} illustrates three such surfaces belonging to an arbitrary object, and having three distinct values of $r_{rc}$. All volume density surfaces share the same shape, orientation, and center position. The only differences between surfaces of different $r_{rc}$ are in their sizes, and volume densities. 
\item[Assumption 2c:] Each volume density surface, when projected onto a plane perpendicular to the line of sight produces a two-dimensional boundary whose area ($A$) is directly proportional to $r_{rc}^2$. The right panel of Figure \ref{cfig1} describes such projections of three surfaces with independent values of $r_{rc}$. All such surface boundaries are identical except in their size and the corresponding volume densities they represent. 
\end{description}

Self-similarity between different volume density surfaces is a critical aspect of assumptions 2b and 2c. Aside from certain constants, the only parameters which differentiate the projected boundaries of different volume density surfaces are their areas (expressable in terms of $r_{rc}^2$), and the corresponding volume density (determined by $f_n(r_{rc})$). Therefore there must be a relationship between the area of each projected boundary and its volume density which is dependent on $f_n(r_{rc})$, but is, aside from some constants, independent of $f_c(\alpha,\theta)$. \emph{$f_c(\alpha,\theta)$ determines the shape and orientation which are identical for each surface, while the relationship between the projected area and volume density of each surface is governed by the radial density profile function ($f_n(r_{rc})$)}. 

The observable column density map is a superposition of all the volume density surfaces projected onto a plane perpendicular to our line of sight. The column density map should thus exhibit the same self-similarity seen among the individual volume density surfaces. If assumptions 1 and 2 hold for a given object, then the following must be valid as well
\begin{description}
\item[Assumption 3:] Comparing the column densities and areas of different column density contours should yield a relationship which, aside from some constants, is independent of the object's geometry. 
\end{description}

Assumption 3 is confirmed analytically by equation \ref{cNarea} in Section \ref{canalyticder}. The following section shows how the function $f_n(r_{rc})$ may be derived using that relationship.

\section{Derivation}
\label{cderivation}
No truly general proof that applies to all possible shapes is evident at this time. Therefore it is necessary to restrict this analytic derivation to those geometries which can be described by a quadratic definition of $r_{rc}$. Geometries which do not conform to equation \ref{crrc} are tested numerically in Section \ref{cnumeric}. A useful form for $r_{rc}$ is

\begin{equation}
\label{crrc}
r_{rc}(x,y,z)=\sqrt{x^2 a + x b(y,z) + c(y,z)} \lc
\end{equation}
	where $a$ is a constant and $b(y,z)$ and $c(y,z)$ are any functions that conform to assumption 2. The above quadratic representation, while not universal, can describe a wide variety of geometries encountered in nature including triaxial ellipsoids. No specific values for $a,b(y,z),$ and $c(y,z)$ are invoked in the following derivation except where noted for purposes of illustration. In such cases, a spheroid model with axial ratio $\alpha$, inclined by an angle $i$ through a rotation about the y axis will be used. A spheroid is chosen because it is mathematically tractable yet versatile enough to demonstrate changes in shape and orientation by varying $\alpha$ and $i$ respectively. Such a spheroid may be described by the relationships
\begin{eqnarray}
\label{cadef}
a=\frac{\omega^2}{a_c^2 \alpha^2} \lc \ \omega^2=\alpha^2 cos^2(i) + sin^2(i) \lc \\
\label{cbdef}
b(z)=\frac{2 z cos(i) sin(i) (1-\alpha^2)}{a_c^2 \alpha^2} \lc \\
\label{ccdef}
c(y,z)=\frac{y^2 \alpha^2 + z^2(\alpha^2 sin^2(i) + cos^2(i))}{a_c^2 \alpha^2} \lp
\end{eqnarray}

Values of $r_{rc}$ describe individual surfaces of fixed volume density. Since the line of sight is chosen to be along the x axis it is useful to express the x positions of each surface with a specific $r_{rc}$ as 
\begin{equation}
x_{r_{rc},\pm}(y,z)=\frac{-b(y,z) \pm \sqrt{b(y,z)^2 - 4 a c(y,z) + 4 a r_{rc}^2}}{2 a} \lp
\end{equation}

Defining a new function
\begin{equation}
E(y,z) = \frac{4 a c(y,z) - b(y,z)^2}{4 a} \lc
\end{equation}
yields
\begin{equation}
\label{cxpos}
x_{r_{rc},\pm}(y,z)=\frac{-b(y,z)}{2 a} \pm \frac{\sqrt{r_{rc}^2 - E(y,z)}}{\sqrt{a}} \lp
\end{equation}

$x_{r_{rc},\pm}(y,z)$ denotes the two line of sight ($x$) positions for a surface defined by a particular value of $r_{rc}$ at sky position $(y,z)$. Functions $a$, $b(y,z)$, and $c(y,z)$ are defined by the geometry of the object in question. Equations \ref{cadef}, \ref{cbdef}, and \ref{ccdef} describe the appropriate functions representing a spheroid with axial ratio $\alpha$ and inclination $i$. $E(y,z)$ is a function that is entirely dependent on the object's shape; the following section discusses its conceptual meaning further.

\subsection{Discrete Derivation}
\label{cdiscretesec}
It is possible to define an object as a discrete series of shells, each of which is defined as the region between an inner ($r_{rc,i}$) and an outer ($r_{rc,o}$) surface with an average volume density ($n_{i,o}$) within the shell. The depth along the x axis of each such shell at different (y,z) positions will vary according to

\begin{equation}
\label{cdori}
D(y,z)_{i,o} = (x_{r_{rc,o},+}(y,z) - x_{r_{rc,o},-}(y,z)) - (x_{r_{rc,i},+}(y,z) - x_{r_{rc,i},-}(y,z)) \lc 
\end{equation}
where $D(y,z)_{i,o}$ is the total depth along the line of sight at position $(y,z)$ for the shell made up of two surfaces defined by $r_{rc,i}$ and $r_{rc,o}$. Substituting equation \ref{cxpos} into equation \ref{cdori} yields
\begin{equation}
\label{cdinc}
D(y,z)_{i,o}= \frac{2}{\sqrt{a}} \left(\sqrt{r_{rc,o}^2 - E(y,z)} - \sqrt{r_{rc,i}^2 - E(y,z)}\right) \lp
\end{equation}
Equation \ref{cdinc} describes the depth of each shell, however this representation is of limited use since $r_{rc}$ is not an observable quantity. Similarly, $E(y,z)$ is a function that is directly dependent on the object's unknown shape. Equation \ref{cdinc} must be put in terms of observable quantities: the observed column density, and the area within each column density contour. 

Each surface as described by equation \ref{cxpos}, when projected onto the line of sight, produces a closed boundary composed of those $(y,z)$ positions where $x_{r_{rc},+}(y,z) = x_{r_{rc},-}(y,z)$. In view of Equation \ref{cxpos} the projected boundaries of each shell are defined by
\begin{equation}
\label{ceboundary}
E(y,z)_{boundary} = r_{rc}^2 \lp
\end{equation}
Solving for $E(y,z)$ using the spheroid model above yields a familiar relation,
\begin{equation}
\label{cespheroid}
E(y,z)_{spheroid} = \frac{1}{a_c^2} \left( \frac{z^2}{\omega^2} + {y^2} \right) \lc
\end{equation}
which is a simple ellipse that results from projecting a three dimensional spheroid surface onto a two dimensional plane. Equation \ref{ceboundary} makes  clear that each contour of equal $E(y,z)$ corresponds to the boundary of a particular volume density surface, with a specific value of $r_{rc}$ and possesses a unique projected area. In general, the projected area of a surface of a particular $r_{rc}$ can be expressed as
\begin{equation}
\label{careaeq}
A=\epsilon r_{rc}^2 \lc
\end{equation} 
where $\epsilon$ is a geometry-dependent unknown constant($\epsilon_{sphere}=\pi a_c^2, \epsilon_{spheroid}=\omega \pi a_c^2$). All positions with equal $E(y,z)$ correspond to the projected boundary of a surface with $r_{rc,c}$ with corresponding area $A_c$. The additional subscript c denotes a specific contour.

Therefore, equation \ref{cdinc} can be reformed in terms of areas as
\begin{equation}
\label{cd}
D_{i,o,c} =\frac{2}{\sqrt{a\epsilon}} \left(\sqrt{A_o - A_c} - \sqrt{A_i - A_c}\right) \lc
\end{equation}
where $D_{i,o,c}$ represents the depth along the line of sight of the shell between surfaces defined by $r_{rc}=r_{rc,i}$ and $r_{rc}=r_{rc,o}$ at all positions defined by the contour formed by the projected boundary of the surface defined by $r_{rc,c}$. The observed column density can then be defined as 
\begin{equation}
\label{cncori}
N_{c}= \sum_{k=c}^{\infty} n_{k,k+1} D_{k,k+1,c} = \sum_{k=c}^{\infty} n_{k,k+1} \frac{2}{\sqrt{a\epsilon}} \left(\sqrt{A_{k+1} - A_c} - \sqrt{A_k - A_c}\right)\lp
\end{equation}
$N_{c}$ represents the column density at all positions $(y,z)$ defined by the projected boundary of the $r_{rc,c}$ surface. $n_{k,k+1}$ represents the mean volume density within the shell whose surfaces are defined by $r_{rc,k}$ and $r_{rc,k+1}$. The column density and area are observable quantities, however $n_{k,k+1}/(a\epsilon)$ are unknowns. The relationship between column density ($N$) and area ($A$) can be obtained through contouring the observed map, yielding a discrete series of contour column densities and associated areas. Using such data it should be possible to obtain information on the quantity $n_{k,k+1}/(a\epsilon)$.

It is useful to define two new variables which will represent the derived volume density profile function.
\begin{eqnarray}
\label{cr'n'}
r'= \sqrt{\frac{A}{\pi}} = \sqrt{\frac{\epsilon}{\pi}} r_{rc} \lc \\
\label{cn'}
n'(r')= n(r_{rc}) \sqrt{\frac{\pi}{a\epsilon}} = n_0 \sqrt{\frac{\pi}{a\epsilon}} f_n\left(r'\sqrt{\frac{\pi}{\epsilon}}\right) \lc
\end{eqnarray}
permitting equation \ref{cncori} to be rewritten as
\begin{equation}
\label{cdiscfinal}
N_{c} = \frac{2}{\sqrt{\pi}} \sum_{k=c}^{\infty} n'_{k,k+1} \left(\sqrt{A_{k+1} - A_c} - \sqrt{A_k - A_c}\right)\lp
\end{equation}

The observed column density map thus yields a series of contours denoted by their column density ($N_{c}$) and area ($A_{c}$). Beginning with the outermost contour with the largest area, and moving recursively inward it is possible to derive a series of $n'_{c,c+1}$ measurements for the object using equation \ref{cdiscfinal}. Equation \ref{cr'n'} yields a series of $r'_{c}$ measurements derived from the contour Areas ($A_c$), yielding $n'(r')$. Equation \ref{cn'} shows that $n'(r')$ is related to $n(r_{rc})$ and $f_n(r_{rc})$ through a series of constants ($\epsilon, n_0, a$) that are all unknown. Knowledge of the object's geometry would yield values for $\epsilon$ and $a$ allowing the determination of $n_0$ and the full definition of the object's radial volume density profile $n(r_{rc})$. Conversely, knowledge of $n_0$ could yield information on the object's geometry. 

Without such a priori knowledge there are limits to the information which may be obtained from a single column density map, however $f_n(r_{rc})$ can be determined to within 2 unknown scalars so as to obtain the form of the volume density profile function. The nature of those two scalars (G and $\chi$) is best elucidated through a non-discrete derivation as discussed in the following two sections. This is done without assuming a specific geometry for the object, or the nature of $f_n(r_{rc})$. This derivation is dependent on obtaining valid $N_c$ vs. $A_c$ measurements from the column density map which may be a non-trivial process when working with real data. Methods for obtaining such measurements are discussed in Section \ref{creal} along with examples of the derivation applied to simulated data.

\subsection{Analytic Derivation using Gaussians and Attenuated Power Laws}
\label{canalyticder}
Equation \ref{cdiscfinal} is useful for deriving $n'(r')$ from real data, and is used in all practical examples in this paper with both simulated and real data. However, it does not necessarily give the most insight into the problem. Any practical application of this theorem requires a strict understanding of the relation between $n'(r')$ and $n(r_{rc})$ with respect to the two scalars which separate them. To this end an analytic derivation is invoked in this section which is equivalent to that in Section \ref{cdiscretesec}, yet is qualitatively different in that it illustrates different aspects of the derived $n'(r')$ function. This derivation does not invoke discreteness, but instead uses integration. The integrals prohibit the use of a truly general form for $f_n(r_{rc})$, thus two radial density profiles are invoked for illustrative purposes along with the same spheroid geometry from Section \ref{cdiscretesec}. A gaussian and an attenuated power law are selected as mathematically tractable profiles that are frequently observed in nature. They may be described as
\begin{equation}
\label{cngnp}
n_g(x,y,z) = n_0 e^{-\frac{r_{rc}(x,y,z)^2}{2}}, n_p(x,y,z)=n_0 (r_{rc}(x,y,z)^2 + 1)^{\frac{-\gamma}{2}} \lc
\end{equation}
where $n_g$ and $n_p$ represent the gaussian and attenuated power law functions respectively. $\gamma$ is a constant greater than 1. This attenuated power law function can be viewed as a form of the well-studied Type IV Pareto distribution. It is inspired by, and represents a more generalized form of the King profile \citep{King62}. The King profile was also used by \cite{King62,Dapp09}, and \cite{Pineda2010} when addressing the problem of density distributions within molecular cloud cores, however they each utilized geometric assumptions which we do not invoke. 

If Assumption 1 holds then the observed column density map for each profile can be written as
\begin{equation}
N_g(y,z)= \int_{-\infty}^{\infty} n_g(x,y,z) dx,\ N_p(y,z)= \int_{-\infty}^{\infty} n_p(x,y,z) dx \lp
\end{equation}

Since specific radial density profile functions are used, it is possible to directly perform each integral, yielding
\begin{equation}
\label{cNdirect}
N_g(y,z)=n_0 \sqrt{\frac{2 \pi}{a}} e^{\frac{-1}{2}E(y,z)},\ N_p(y,z)=n_0 \sqrt{\frac{2\pi}{a}} \sqrt{2\pi} {\gamma-2 \choose \frac{\gamma-2}{2}} \left(E(y,z) +1\right)^{\frac{-\gamma+1}{2}} \lc
\end{equation}
where ${\gamma-2 \choose \frac{\gamma-2}{2}}$ is the binomial coefficient. 
Similarly to Section \ref{cdiscretesec}, the preceding equation may be used to express the column density in terms of the area covered by each column density contour resulting in
\begin{equation}
\label{cNarea}
N_g(A)=n_0 \sqrt{\frac{2 \pi}{a}} e^{\frac{-A}{2\epsilon}},\ N_p(A)=n_0 \sqrt{\frac{2\pi}{a}} \sqrt{2\pi} {\gamma-2 \choose \frac{\gamma-2}{2}} \left(\frac{A}{\epsilon} +1\right)^{\frac{-\gamma+1}{2}} \lp
\end{equation}
Since no specific shape has yet been invoked, Equation \ref{cNarea} verifies Assumption 3 by showing that the relationship between column density and the area of its contours is, aside from some constants ($\epsilon, a, n_0$), independent of the object's geometry.

Alternatively, using equations \ref{cNdirect} and \ref{cxpos} along with the relation
\begin{equation}
\frac{d x_{r_{rc}}}{dr_{rc}} = \frac{r_{rc}}{\sqrt{a(r_{rc}^2 - E(y,z))}}
\end{equation}
yields the following expression for the column density which is equivalent to the derivation in section \ref{cdiscretesec}
\begin{equation}
\label{cNrrc}
N(r_{rc}) =  \int_{r_{rc}}^{\infty} n(r_{rc,e}) \frac{2 r_{rc,e}}{\sqrt{a(r_{rc,e}^2 - r_{rc}^2)}} dr_{rc,e} \lp
\end{equation}
Equation \ref{cNrrc} specifies the observed column density at all positions $(y,z)$ described by the projected boundary of the shell defined by $r_{rc}$. $r_{rc,e}$ is the integration variable, where the e subscript denotes that the integration is performed over all surfaces exterior to $r_{rc}$. Solving equation \ref{cNrrc} for the gaussian and attenuated power law profiles yields
\begin{equation}
\label{cNrrc2}
N_g(r_{rc}) = n_0 \sqrt{\frac{2 \pi}{a}} e^{-\frac{r_{rc}^2}{2}}, N_p(r_{rc})=n_0 \sqrt{\frac{2\pi}{a}} \sqrt{2\pi} {\gamma-2 \choose \frac{\gamma-2}{2}} \left(r_{rc}^2 +1\right)^{\frac{-\gamma+1}{2}} \lp
\end{equation}
Equation \ref{cNrrc2}, when converted to Areas using Equation \ref{careaeq}, is identical to Equation \ref{cNarea}, thus confirming that the discrete derivation in Section \ref{cdiscretesec} is equivalent to integrating the volume density along the line of sight.

\subsection{Understanding $n'(r')$}

Deriving $n'(r')$ through the method defined above yields a function with the same form as the original $n(r_{rc})$. It is important to understand the relation between the derived and actual density profile functions. This relationship may be defined as

\begin{equation}
\label{cGepsilon}
		n'(r')=G n(r_{rc}) = G n\left(\frac{r'}{\chi}\right) \lc
\end{equation}
where $G$ and $\chi$ are unknown constants. Applying the above method to the spheroid model with gaussian and power-law profiles yields derived volume density functions described by

\begin{equation}
\label{cnprimegnprimep}
	n'_g (r') = G n_0 e^{\frac{-r_{rc}^2}{2 \chi^2}},\ n'_p(r') = G n_0 \left( \frac{r_{rc}^2}{\chi^2} + 1 \right) ^{\frac{-\gamma}{2}} \lp
\end{equation}
Equations \ref{cr'n'} and \ref{cGepsilon} in conjunction with Equation \ref{cnprimegnprimep} show that for a spheroid model, 
\begin{equation}
\label{cGchi}
G = \sqrt{\frac{\pi}{a \epsilon}} = \frac{\alpha}{\omega^{\frac{3}{2}}}, \chi = \sqrt{\frac{\epsilon}{\pi}} =\sqrt{\omega} a_c \lp
\end{equation}
$G$ is dimensionless, while $\chi$ has dimensions of length. It is important to note that $G$ and $\chi$ are completely geometry dependent and thus identical in both the power law and gaussian cases. Neither parameter can be fully determined by the method described here without knowledge of the object's geometry, further data, or assumptions. As evidenced by Equation \ref{cGchi}, $G$ and $\chi$ are not independent quantities  due to their dependence on $\omega$. Aside from scalars $G$ and $\chi$, the derived $n'(r')$, and the original $n(r_{rc})$ are identical. For a sphere, $G = 1$ and $\chi = a_c$. These scalars contain all of the unknown geometric information about the observed object. We may derive $n'(r')$ from an observed column density map, however this function will differ from the object's volume density profile ($n(r_{rc})$) by the two unknown scalars $G$ and $\chi$. The form of $n'(r')$ will however be identical to $n(r_{rc})$ regardless of the two scalars. If an object's geometry is known, $G$ and $\chi$ may be calculated (values for the spheroid are shown in Equation \ref{cGchi}). In the most general terms, $G$ may be viewed as the ratio between the depth and the width of an object along the line of sight, though this interpretation will rarely be strictly true. If $G$ is greater than 1, then the object is deeper than it is wide, and the derived $n'_0$ will be greater than the actual $n_0$. 

\section{Examples using simulated data}
\label{cnumeric}

	Numerical models using simulated data can validate the technique described in Section \ref{cderivation}, as well as illustrate the behaviour of $G$ and $\chi$ under various conditions. Models of several objects are constructed using known geometries and volume density profiles in order to create simulated column density maps. These maps are then used to derive $n'(r')$ which are finally compared to the original (known) volume density functions used to construct the model.

\subsection{Model Construction and Analysis}

\begin{figure}
\includegraphics[scale=.75]{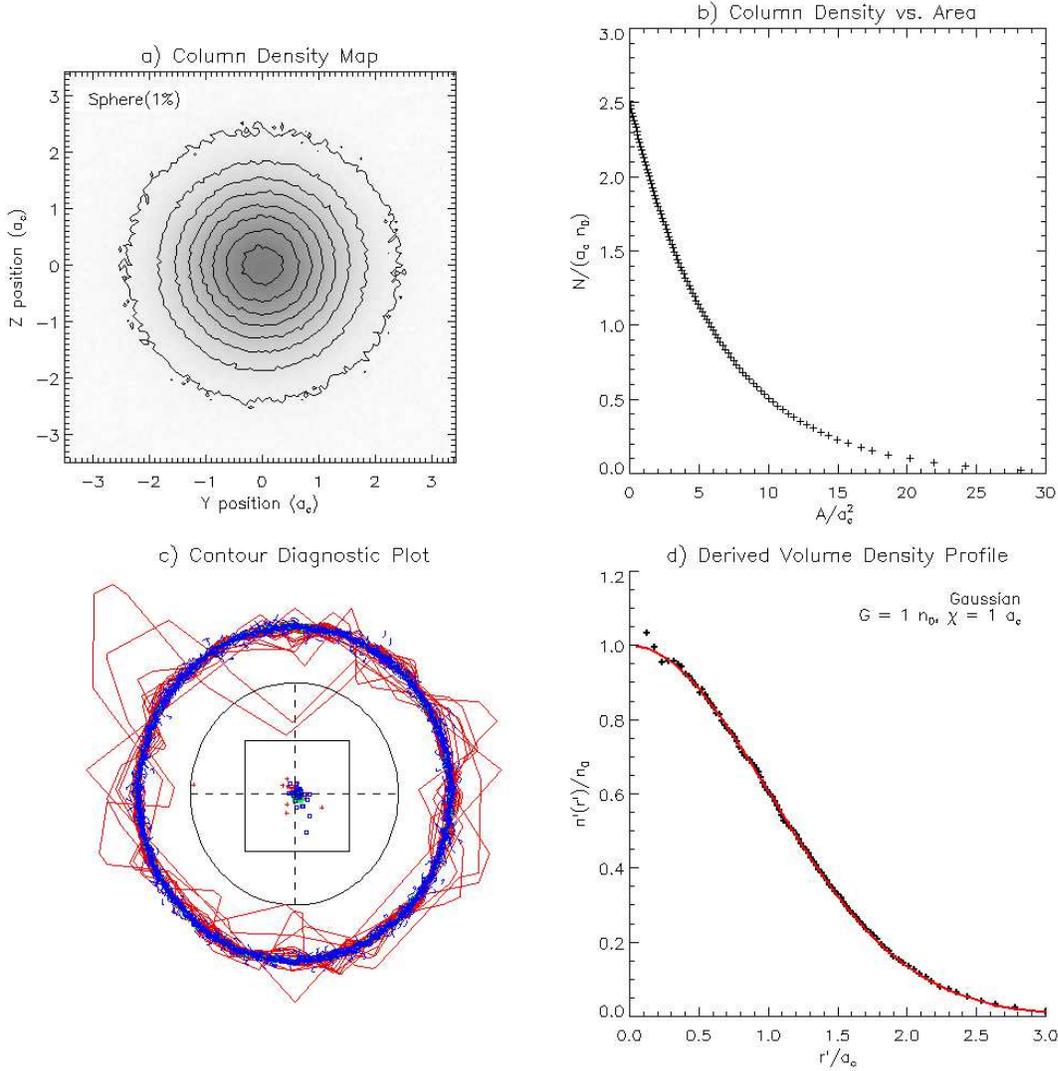}
\caption{A model object with spherical geometry ($a=1/a_c^2, b=0, c=(y^2 + z^2)/a_c^2$), and a radial volume density profile described by a gaussian ($n(r_{rc}) = n_0 e^{-r_{rc}^2/2} $). a) A simulated column density map with sample contours. Gaussian noise is added equivalent to 1\% of the maximum column density. $Y$ and $Z$ coordinates are represented in units of $a_c$. b) Column Density ($N$) and corresponding Area ($A$) for each contour (not displayed) used in the analysis. c) A contour diagnostic plot for the object, as described in Section \ref{cnumeric}. d) The derived volume density profile function ($n'(r')$). Black points represent the values derived from each $N$ and $A$ contour pair. The red line represents the original function $n(r_{rc})$ used by the model as scaled by $G$ and $\chi$.     
\label{cfig2}}
\end{figure}

	Figure \ref{cfig2} illustrates how such a model is constructed and analyzed using the simplest case of a sphere with a gaussian volume density profile and minimal noise. To produce a column density map such as in Figure \ref{cfig2}a it is necessary to first choose a geometry (in this case a sphere) and a volume density profile function (in this case a gaussian) and construct a three-dimensional array whose elements represent the object's volume density. This array is then integrated along the line of sight to produce a column density map. Normally distributed noise with mean zero and a certain standard deviation (1\% of the maximum column density in the case of Figure \ref{cfig2}a) is then added to the column density map. Selected contour levels are drawn for illustration purposes to produce a map as in Figure \ref{cfig2}a. Such column density maps are the only source of information for further model analysis, as knowledge of model scalars such as $a_c$ and $n_0$ is used only for the purposes of scaling the plots. 
	
	Many column density contours (not drawn) are measured on the map in order to produce a plot of column density ($N$) versus contour area ($A$) as in Figure \ref{cfig2}b. It is impossible to properly sample the whole range of column densities without a priori knowledge of the volume density profile function. We found it most appropriate to measure the same number of contours as the number of pixels that span the object, and to space them equally in column density. This choice often results in oversampling, as discussed in Section \ref{cuncertainty}, but has been experimentally found to be the most useful.

	The technique described here requires implicitly that all column density contours exhibit self-similarity, sharing the same shape, orientation, and center position. 
The suitability of an object to this analysis technique may be verified by comparing the contours. 
Similarly, it is necessary to remove from consideration any contours which are created by noise in the column density map. 
Figure \ref{cfig2}c illustrates how these requirements are satisfied. Each contour is scaled to the same size, translated to the same center position, and then plotted so as to overlap as in Figure \ref{cfig2}c. 
The innermost third of the contours with the smallest area are colored red, while the outermost third are colored blue, and intermediate contours are colored green. 
In the case of Figure \ref{cfig2}c the simulated noise is quite low and thus only the innermost (smallest) contours display any deviations from a circle. 
These variations are due to the small number of pixels within the smallest contours. 
This representation is useful in that any deviations from a single contour shape and orientation may be easily identified.

	A simple method for numerically filtering out noise-induced contours from consideration is to compare the geometric centers of each contour to the geometric center of mass for the object from the column density map. Any contours whose centers exceed some small distance from the center of mass are excluded. Figure \ref{cfig2}c represents the object's center of mass as the dashed cross in the center. The center positions of each contour are plotted in relation to the center of mass with the green, red, and blue colors representing the contours with the smallest, intermediate, and largest areas respectively. The solid black circle represents the radius used to filter out questionable contours. The square represents the relative position and scale of the pixel from the column density map which contains the object's center of mass. This diagnostic plot is useful in determining how well a given object complies with assumption 2, as well as which contours are suitable for analysis.   
 
	Once all unsuitable column density contours are removed from consideration it is possible to apply Equations \ref{cr'n'} and \ref{cdiscfinal} recursively to the $N$ and $A$ pairs in order to derive $n'(r')$ as in Figure \ref{cfig2}d. Since modeled data is used here it is possible to directly determine the values of $G$ and $\chi$ as well as to scale $r'$ and $n'(r')$ using the known values of $a_c$ and $n_0$ as in Figure \ref{cfig2}d. 
	
\subsection{Tests Using Various Geometries and Profile Functions}	
\label{csimulatedgeometries}	
\begin{figure}
\includegraphics[scale=.75]{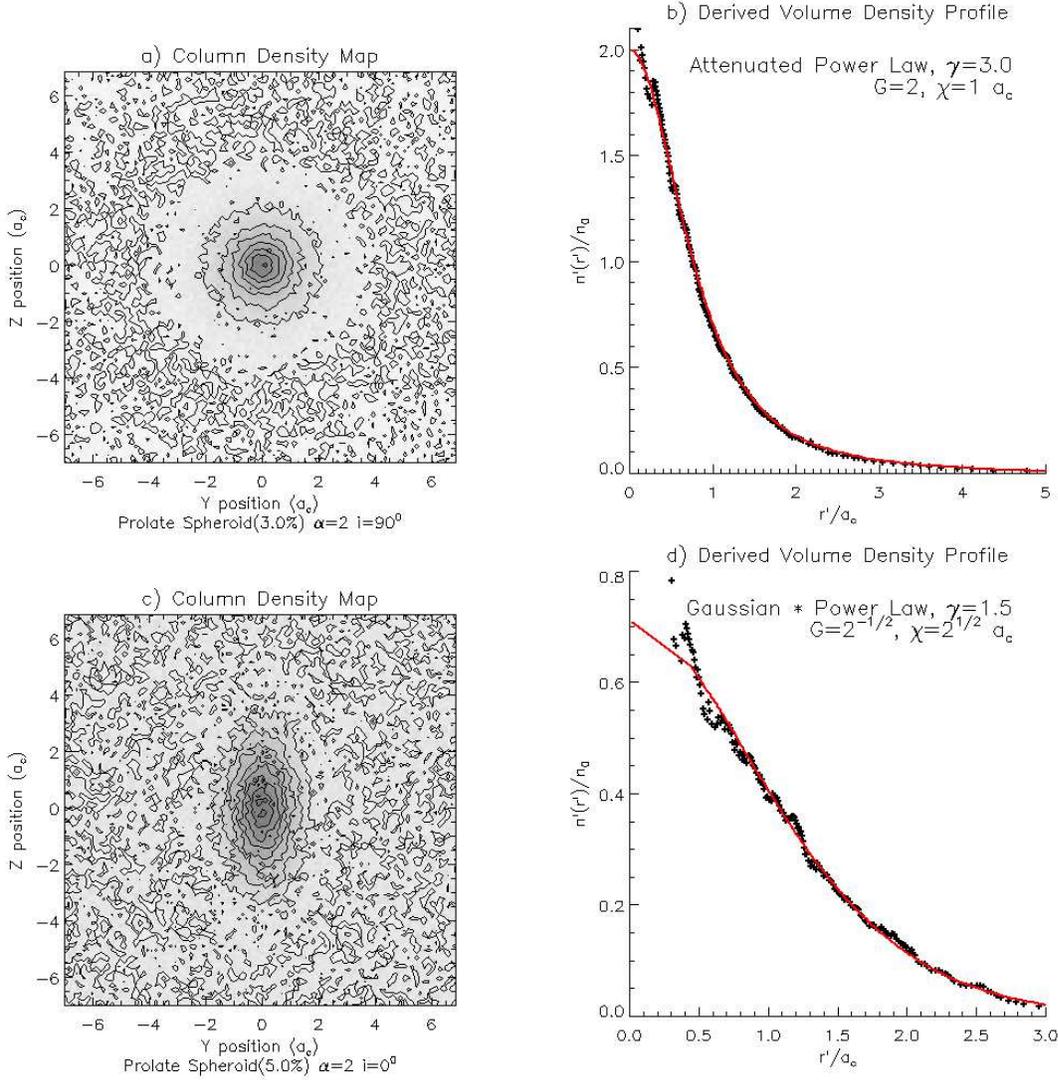}
\caption{a) Simulated column density map of a prolate spheroid as described by Equations \ref{cadef}, \ref{cbdef}, and \ref{ccdef} with $\alpha = 2$, $i = 0^o$, and 3\% noise added. b) Actual (red line) and derived (black dots) volume density profile for the object in a. An attenuated power-law as in Equation \ref{cngnp} with $\gamma = 3$ is used to construct the object in a and b. c) Simulated column density map of an object using the same geometry as in a, except that the object is rotated by $90^o$ about the y axis and 5\% noise is added. d) Actual (red line) and derived (black dots) volume density profile for the object in c. The radial volume density profile used in c and d is given by $n(r_{rc})= n_0 e^{-r_{rc}^2/2} (r_{rc}^2 + 1)^{\frac{-\gamma}{2}}$ with $\gamma = 1.5$.       
\label{cfig3}}
\end{figure}

\begin{figure}
\includegraphics[scale=.75]{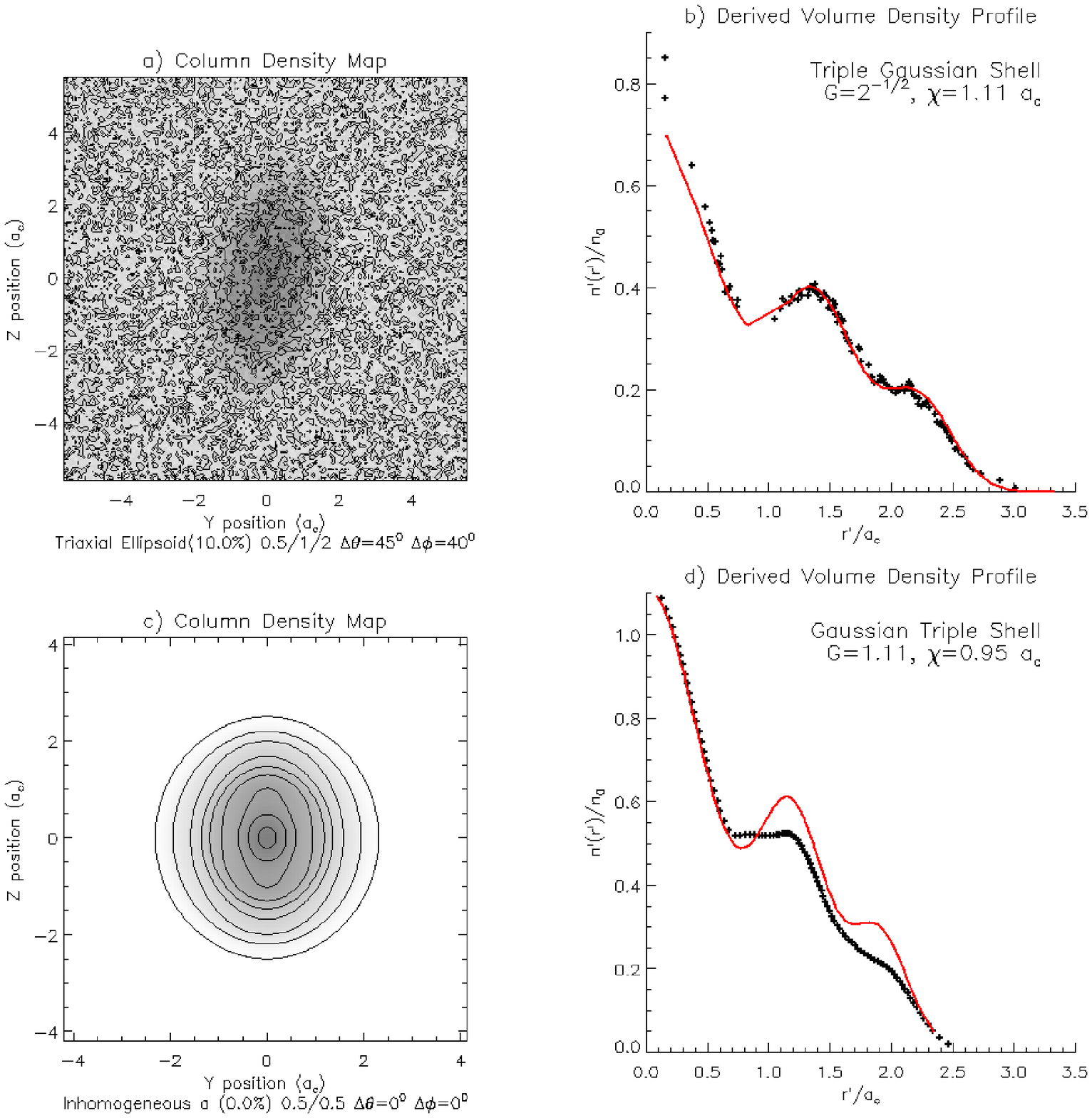}
\caption{a) Simulated column density map of a triaxial ellipsoid with axis dimensions of 0.5,1, and 2 $a_c$ and 10\% noise added. b) Actual (red line) and derived (black dots) volume density profile for the object in a. A triple gaussian volume density profile function is used to construct the object in a and b. c) Simulated column density map of an object with nonuniform $a$ (Equation \ref{cadef}). d) Actual (red line) and derived (black dots) volume density profile for the object in c. Since contour self-similarity is not present throughout, the object is not expected to be adequately modelled by this technique. The radial volume density profile used in c and d is described by the same triple gaussian as in a and b.       
\label{cfig4}}
\end{figure}

	Figure \ref{cfig2} verifies that the derived $n'(r')$ has the same form as the original $n(r_{rc})$ function to a very high degree for the low-noise sphere with a gaussian profile function. 
As expected, $G=1$ and $\chi = a_c$.
The technique described in Section \ref{cderivation} should apply to any profile function, as well as to any geometries which fulfill the requirements described in Section \ref{cassumptions}. 
To that end, Figures \ref{cfig3}a-b, \ref{cfig3}c-d, and \ref{cfig4}a-b present three cases beyond the simple gaussian sphere in Figure \ref{cfig2}. 
Figures \ref{cfig3}a-b, and \ref{cfig3}c-d represent the spheroid defined in Section \ref{cderivation} with two different orientations, while Figure \ref{cfig4}a-b represents a tri-axial ellipsoid. 
The derived and original volume density profile function forms agree to a great extent, verifying that the technique is valid for tri-axial ellipsoids of any shape and orientation. 
Several other geometries (not shown) which satisfy the assumptions in Section \ref{cassumptions} were tested and were all shown to produce valid results. 

	The derivation in Section \ref{cderivation} requires that $a$ be a constant (Equation \ref{crrc}). 
Any geometry which involves a definition of $a$ that is dependent on spatial coordinates $y$ and $z$ results in an object which does not conform to the assumptions in Section \ref{cassumptions}, and thus is not suitable to the analysis presented here. 
This is due to the variation in the depth of each shell resulting from an inhomogeneous $a(y,z)$ as in Equation \ref{cd}.
If the object has a non-constant value of $a(y,z)$ then the relationship between the area and depth of each shell is no longer a constant, and the technique described by Equation \ref{cncori} fails as $a$ would be dependent on the shell number $k$. 

	It is possible to determine from the column density map whether the object in question has a geometry which is dependent on a constant value of $a$. 
Equation \ref{careaeq} shows that the projected area of each shell is dependent on $\epsilon$. 
From the definition of the spheroid it can be shown that $\epsilon_{spheroid}=\omega \pi a_c^2 = \sqrt{a} a_c \alpha \pi a_c^2$. 
If $a$ is nonuniform then so is $\epsilon$, meaning that the relationship between a shell's area and $r_{rc}^2$ is no longer constant. 
This implies that the projected boundary of each shell, and thus each column density contour, has a different shape.
\emph{Inhomogeneities in contour shape invalidate this technique}. 
Figures \ref{cfig4}c-d show such a geometric shape which utilizes the same quadratic definition for $r_{rc}$ as in Equation \ref{crrc} where
\begin{eqnarray}
\beta=0.5, \gamma=0.5 \lc \\
a(y) = 1 + \gamma y^2 \lc \\
b(y) = -2 \beta^2 \gamma y^2 \lc \\
c(y,z) = (1 + \beta^2 \gamma) y^2 + z^2 \lp
\end{eqnarray}

In this case, $a$ is not a constant, the object does not produce self-similar contours, and our technique fails to reproduce the correct radial profile as seen in Figure \ref{cfig4}d.
It is important to note however, that the inner-most and outer-most regions (where there contours are self-similar) do correctly reproduce the original profile function.
Fortunately, the influence of each individual contour of the overall density profile is limited. 
Each contour only affects contours interior to itself, and has the greatest influence on adjacent contours. 
Thus, in our derivation, as one moves from the outermost shells to the innermost, a change in the value of $\epsilon$ will only begin to have an effect at the contour where the change first occurs, and its influence will decline as we move further into the interior. 
We refer to this change in contour shape as an $\epsilon$ discontinuity. 
Assuming only one such discontinuity occurs within a given map (so that there are only two contour shapes present), the derived volume density profile should still be correct in the region outside of the discontinuity.
The discontinuity will invalidate the derived profile interior to itself, but as its influence weakens the innermost region of the derived profile may still be accurate, as in Figure \ref{cfig4}d.
This phenomenon is frequently present when dealing with real data, and is illustrated further in Section \ref{creal}, but it only prohibits us from accurately obtaining $n'(r')$ in those portions of the profile where $\epsilon$ discontinuities occur. $\epsilon$ discontinuities are simple to identify visually through plots such as Figure 2c, and numerically by calculating the values of $\epsilon$ for each contour.

\subsection{Understanding Uncertainties and Distinguishing Real Features From Noise and Systematic Effects}
\label{cuncertainty}
Sections \ref{cderivation} and \ref{cnumeric} have proven that our technique can accurately derive the form of the radial volume density profile under idealized conditions. 
Such circumstances rarely occur in nature so it is important to be able to distinguish real data from noise and systematic effects.

Ideally, one would assign an uncertainty to each measurement in the derived profile to determine which points are likely to represent real measurements. 
However, a serious shortcoming of our method here is that we are unable to assign such uncertainties. 
Many methods were tried, but the root problem is that we are unable to properly assign an accurate uncertainty to the $N$ vs. $A$ measurements for the contours. 
Even assuming that all contours ultimately have the same shape, systematic noise affects them in ways that are difficult to quantify. 
Each of the above simulated data sets, are made with different levels of systematic noise in order to illustrate its effects on the solutions. 
Figure \ref{cfig2} has minimal noise equivalent to $1\%$ of the peak intensity, resulting in a near perfect derived profile. 
Figures \ref{cfig3}a-b contain $3\%$ noise which is sufficient to add some irregularity to the shapes of the observed contours. 
Those irregularities are most evident in the smallest contours, and are seen as slight offsets in the innermost regions of the radial density profile. 

The $5\%$ noise map in Figures \ref{cfig3}c-d shows a new phenomenon. 
Here, the noise is such that the innermost contour is broken into two pieces which are unusable. 
As a result, the derived profile has no measurements interior to $r'= 0.3 a_0$.
Further, there are larger irregularities in the derived profile. 
These are caused by individual pixels with a significantly different signal compared to their surroundings. 
Contours will tend to bend around such pixels, until a certain column density threshold is reached and the contours snap onto the other side of the pixel. 
This phenomenon may be recognized in that all the irregularities will have roughly the same width in the derived profile, corresponding roughly to the width of each pixel in the map as evident in Figure \ref{cfig3}d.  
Figures \ref{cfig4}a-b show a more difficult case in which the noise is $10\%$ of the peak signal. 
The object in the map can be difficult to discern under such conditions. 
The systematic noise will prevent the construction of the innermost contours, but may also prevent the formation of contours in other regions leading to gaps in the derived profile such as in Figure \ref{cfig4}b. It is important to note that the gaps did not prevent the derivation of the correct profile in the regions where contours could be formed.

Changes in contour shape introduce a bias to the resulting profiles, as seen in Figure \ref{cfig4}c-d. 
No reliable method for removing such bias is apparent. 
However, the bias seems to be localized to only those regions of the derived profile immediately interior to the $\epsilon$ discontinuity. As a result, the innermost portion of the derived density profile in Figure \ref{cfig4}d agrees well with the original profile. How these $\epsilon$ discontinuities manifest themselves is best illustrated with real data as seen in the following section.

\section{Derivation of Molecular Cloud Core Volume Density Profiles Using Real Data}
\label{creal}

Star formation theory abounds with open questions, many of which are related to the process by which dense cores within molecular clouds collapse. 
Of particular interest is the balance between forces which induce collapse, such as gravity and external pressure, and support mechanisms such as thermal pressure, turbulence, angular momentum, magnetic fields, etc. 
A cloud core's density distribution is central to understanding this balance.
Thus measuring both the gas and dust components to obtain the distribution of the total proton density is of critical importance, representing a long-standing and active field of study.

The basic model of an equilibrium mass distribution \citep{Bonnor1956, Ebert1955} assumes an isothermal sphere bounded by some external pressure. 
Even such a simple model yields powerful insights, such as that radial volume density profiles in molecular cloud cores should resemble power laws with an approximate $n\ \alpha \ r^{-2}$ relationship between radius and volume density at the edge of the core, with a weaker dependence on radius towards the center of the core.
Several recent studies have utilized Bonnor-Ebert spheres, or some derivative thereof, such as \cite{Evans2001}, \cite{Alves2001a}, and \cite{Teixeira2005}. 
As mentioned above, some studies \citep[e.g.][]{Cernicharo1985, Arquilla1985} have fitted a power law to the radial density profile.  While often satisfactory, there is considerable variation in the value of the exponent among different cloud cores, with $n \propto r^{-1~{\rm to}~-3}$ found in these studies.  

Precise measurements of the exponents in cloud cores speak to the significance of the support mechanisms. 
Studies employing a variety of techniques have convincingly demonstrated that clouds in different evolutionary states exhibit different density distributions \citep[with selected examples being][] {Ward-Thompson94, Kainulainen07, Liu2012, Wu2012}. 
\cite{Liu2012} made a strong case for the urgent need of investigations of density distributions and support mechanisms in pre-stellar cores in light of new data from Planck.

We have chosen to demonstrate our technique using total proton counts in molecular clouds due to their well-known power-law nature which may be used to validate the technique. 
By not assuming a geometry, we remove a significant source of bias present in previous observations. 
The relatively small sample size here is used for demonstration purposes, while a more focused study will be presented in forthcoming publications. 

It is assumed under most circumstances that gas and dust are fairly well mixed in the diffuse ISM and in molecular clouds. 
Some molecular species, such as $H_{2}$ or $^{13}$CO are generally good tracers of the total proton count in molecular clouds. However, $H_{2}$ cannot be directly measured in clouds unless through absorption against a background source. 
$^{13}$CO requires us to determine its excitation temperature to obtain column densities.
Further, carbon monoxide has been shown to freeze onto dust grains at higher densities \citep{Kramer1999, Tafalla2002, Bergin2002, Pineda2010}. 
While not without its limitations, dust provides a well-tested, proven alternative.
Estimating total proton column densities through stellar reddening avoids the need to determine temperatures and may be used ubiquitously throughout a cloud assuming sufficient background stars are visible. 
\cite{Goodman09} compared dust extinction, Near-Infrared Emission, and $^{13}$CO emission as probes of the total proton content of dense clouds.
After a detailed examination they concluded that dust extinction provided the simplest and most reliable probe.
Methods for deriving total proton column densities through stellar reddening data are well--developed \citep[e.g.][]{Lada94, Lombardi01,Dobashi11} and can be readily employed using infrared data from the 2MASS sky survey \citep{2MASS}. 
The 2MASS data in conjunction with these methods provide a widely accessible, and comparatively uncontroversial method for obtaining total proton column density maps.

The data for two of the sources, which might actually be described as either dense clouds or cloud cores, (B133, and L466) were derived by the authors from the raw 2MASS stellar reddening catalog using an implementation of the NICER method \citep{Lombardi01, Chapman2009}. 
Maps for the other clouds (L1765, L1709, B5, NGC1333) were obtained from the Perseus and Ophiuchus final extinction maps as part of the COMPLETE survey that also utilize the NICER method \citep{Lombardi01}. 
L1765, and L1709 are part of the Ophiuchus Complex, while B5 and NGC1333 are part of the Perseus complex. Each of these four clouds is in a region with substantial background extinction and is accompanied by neighboring features.  For these, a 3 arcminute beam size was used. B133 and L466 are more isolated, with comparatively little background extinction or neighboring features. The maps for B133 and L466 employ a 1 arcminute beam.

\begin{figure}
\includegraphics[scale=1.0]{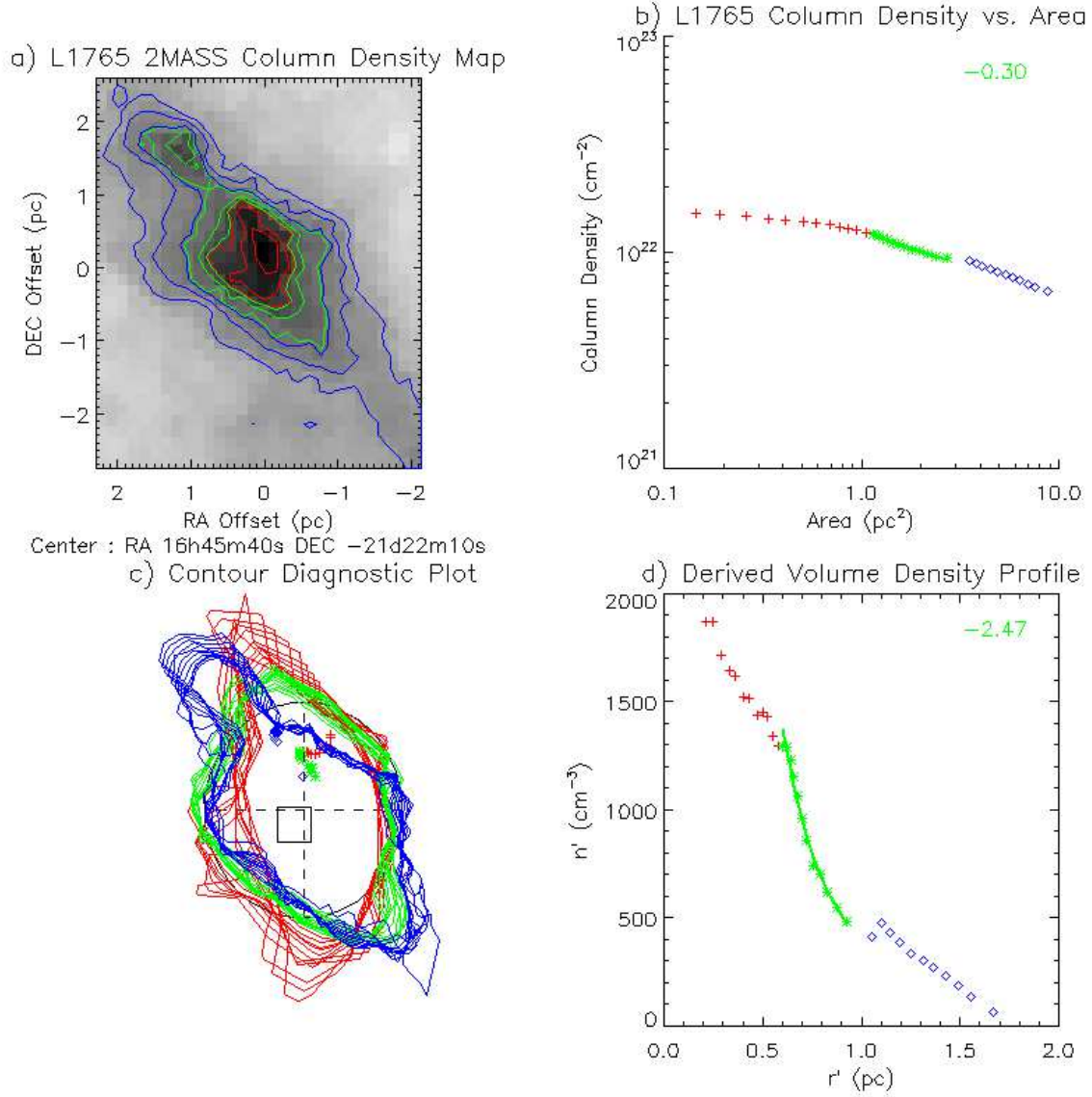}
\caption{a) L1756 Column Density map. Darker pixels represent greater column density. Sample contours are drawn, with each color representing an individual contour group. b) $N$ vs. $A$ plot derived from contours actually used in the derivation, color groups correspond to the coloring in a. The fitted line(s) correspond to simple power-law fits for each color-group with the exponent(s) printed in the legend, and are only drawn through segments which are believed to be trustworthy. c) Contour Diagnostic Plot similar to that in Figure \ref{cfig2}a. d) Derived volume density profile. The fitted line(s) correspond to simple power-law fits for each color-group with the exponent(s) printed in the legend, and are only drawn through segments which are believed to be trustworthy.       
\label{cfig5}}
\end{figure}

The scheme used in Figure \ref{cfig5} is used to describe each of the clouds in this study.
Figure \ref{cfig5}a represents the column density map for L1756, with some sample contours added. 
Column density contours are applied throughout the map and filtered as described in Section \ref{cnumeric} to produce the column density vs. area ($N$ vs. $A$)plot in Figure \ref{cfig5}b. 
It is apparent that there are three different behaviors in the $N$ vs. $A$ plot, therefore each measurement has been coded with a symbol and color. 
This color and symbol scheme is applied throughout the whole Figure.
The colors of the contours drawn on the column density map correspond to the same column density levels as the colors in the $N$ vs. $A$ plot.
Figure \ref{cfig5}c is a diagnostic plot similar to Figure \ref{cfig2}c. 
The derived volume density profile is depicted in Figure \ref{cfig5}d. 
In contrast to the simulated data in Figure \ref{cfig2}d, $n_0$ and $a_0$ for the real cloud are not known, and thus the plot is scaled in terms of $n'$ and $r'$. 

Three regions are evident in L1765 with red crosses representing the highest column density contours, blue representing the outer-most contours, and green those in between. 
The changes in behavior between the three groups are in fact characterized by two $\epsilon$ discontinuities. 
In the case of L1765, only the green data are believed to be trustworthy. 
The red group represents the inner core. In the column density map and the contour diagnostic plot, the red contours do not appear to be self-similar. 
The contours may in fact be self-similar, but there are too few pixels to properly define their shape, and thus they are not reliable. 
Experimentation with real data reveals that the smallest contours must have an area greater than approximately 25 Nyquist-sampled pixels to be sufficiently well-defined and thus be usable.
With L1765, these innermost contours are displayed as an example; they are removed from consideration in the other clouds. The map shows green contours centered around the main cloud, as well as separate contours along the secondary clump. 
The blue contours however encircle the secondary clump as well. 
This technique cannot correctly function where there are two cores. 
As a result the blue contours and the related data cannot be trusted.
Only the green contours around the main cloud are trusted.

The green group in L1765 seems to exhibit a very good power law with slope $-0.3$ in the $N$ vs. $A$ plot, and the fitted line is drawn in green in Figure \ref{cfig5}b. 
From Section \ref{cderivation} it is expected that if the $N$ vs. $A$ exhibits a power-law behavior, then so should the derived volume density profile. 
Figure \ref{cfig5}d indeed shows that the green region's volume density profile function follows a power law with slope $-2.47$. 

We are thus able to determine the form of the volume density profile in the intermediate region of the cloud (the green group), where the contours are well-defined, self-similar, and include only one core or clump. 
A reasonable concern is how can the green region be trusted when the red and blue are not. The red measurements are interior to the green ones, and thus have no effect at all on the green measurements. 
Equation \ref{cd} reveals that the depth of each shell is roughly constant in the region interior to the shell, and thus so is its contribution to all interior shells in the derived profile. 
The primary contribution of the blue measurements is to add an approximately constant volume density to the interior green and red measurements during the derivation process. 
However, that constant contribution is irrelevant unless we know the specific geometry of the cloud. 
Only the changes in the shell depth near the edge of each shell can alter the form of the derived interior profile. 
Therefore, only the few inner-most points in the derived profile interior to the $\epsilon$ discontinuity are affected. 

Our derived profile yields values for $n'$ and $r'$ that are scaled by unknown geometry-dependent constants. 
Based on Figure \ref{cfig5}d, we cannot say that at a radius of $0.9$ pc, the volume density within L1765 is equal to $500\ cm^{-3}$, nor is that the goal of this research. 
We can however say with significant confidence that in those regions of the main cloud encompassed by the green contours in Fig \ref{cfig5}a, or approximately the middle third of the cloud radially, the volume density profile is governed by a power law with exponent $-2.47$. 
If, and only if, the cloud is assumed to be approximately spherical, $n'$ and $r'$ may actually represent similar values to the real $n$ and $r$. 
Without geometry information however, we can only be certain of the profile's form within the trusted region.

The fact that we observed a strict power-law within a molecular cloud is in line with previous observations made through other methods. 
It is both encouraging and disconcerting that the derived profile follows a power law quite so well. It is encouraging to see that a real map, with real data, will produce an orderly volume density profile function (in the green region) and that a power-law is observed as it has been by with previous studies. However, it is necessary to make certain that the power-law is not a systematic effect of our technique or of the data itself. We therefore examine additional clouds to determine their behavior, and verify the validity of our technique.

\begin{figure}
\includegraphics[scale=1.]{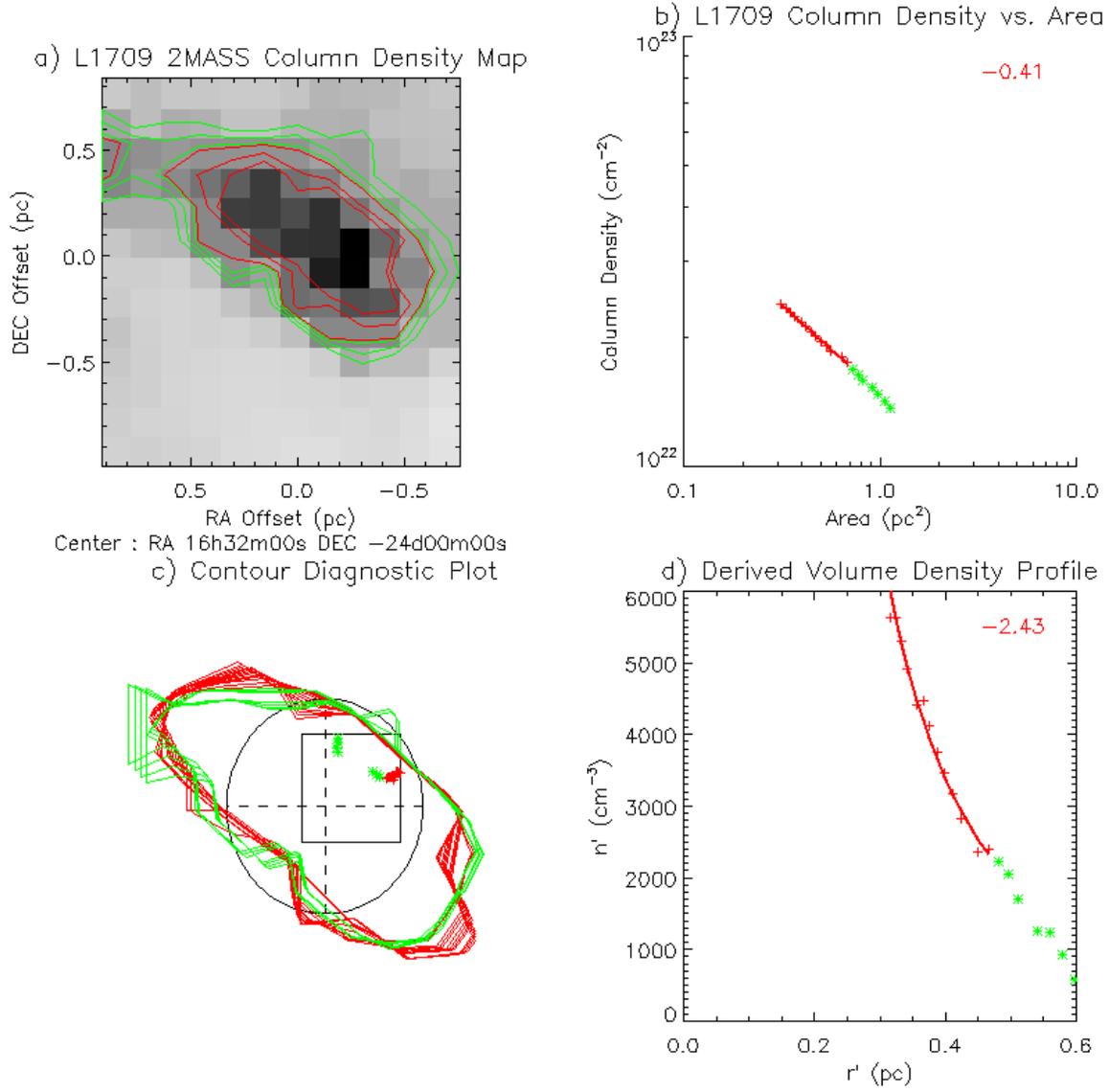}
\caption{L1709 analyzed as in Figure \ref{cfig5}
\label{cfig6}}
\end{figure}

Figure \ref{cfig6} represents L1709. In this case, the region interior to the red group is not shown as those contours have too small an area to be useful. The green contours however, are not trustworthy as they are influenced by the secondary peak at the edge of the map. 
The contours for the red group in Figure \ref{cfig6}c exhibit a remarkable self-similarity in shape and center position even though they vary in area from 0.3 to 0.7 square parsecs. 
Similarly to L1765, the derived profile exhibits a strong power-law with a slope of $-2.43$. 

\begin{figure}
\includegraphics[scale=1.]{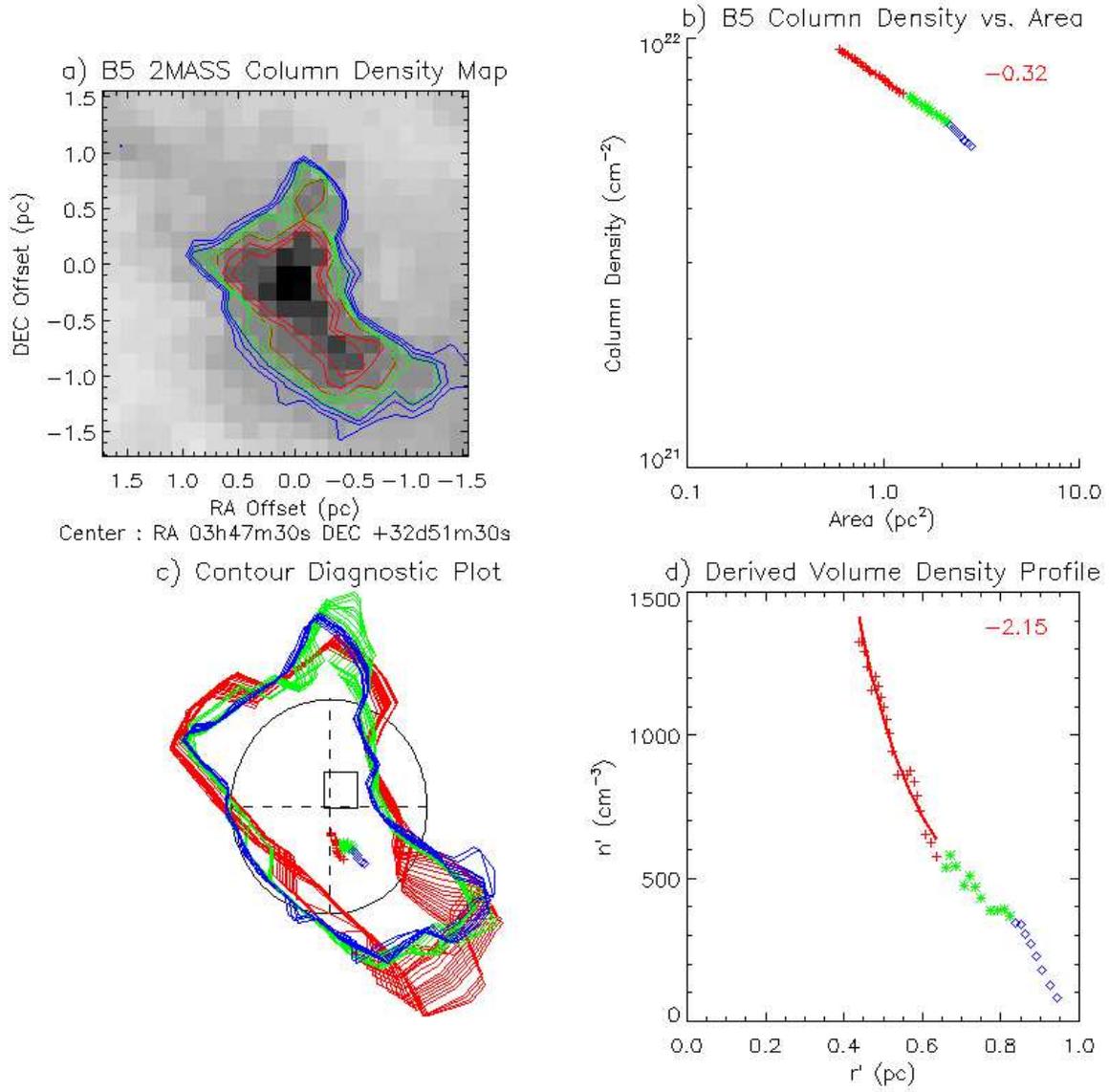}
\caption{B5 analyzed as in Figure \ref{cfig5}
\label{cfig7}}
\end{figure}

The analysis for B5 is depicted in Figure \ref{cfig7}. The blue and green regions are not trustworthy in this case because they encompass two secondary clumps in the top, and bottom right regions of the map. The red contours do not exhibit quite the same level of self-similarity as found in L1756 and L1709 due to the distension in the bottom right region. As a result, even the red contours may be somewhat suspect, however the distension corresponds to a variance of less than $5\%$ in the value of $\epsilon$ among the red contours. The red region corresponds to a power law with slope of $-2.15$.

\begin{figure}
\includegraphics[scale=0.65]{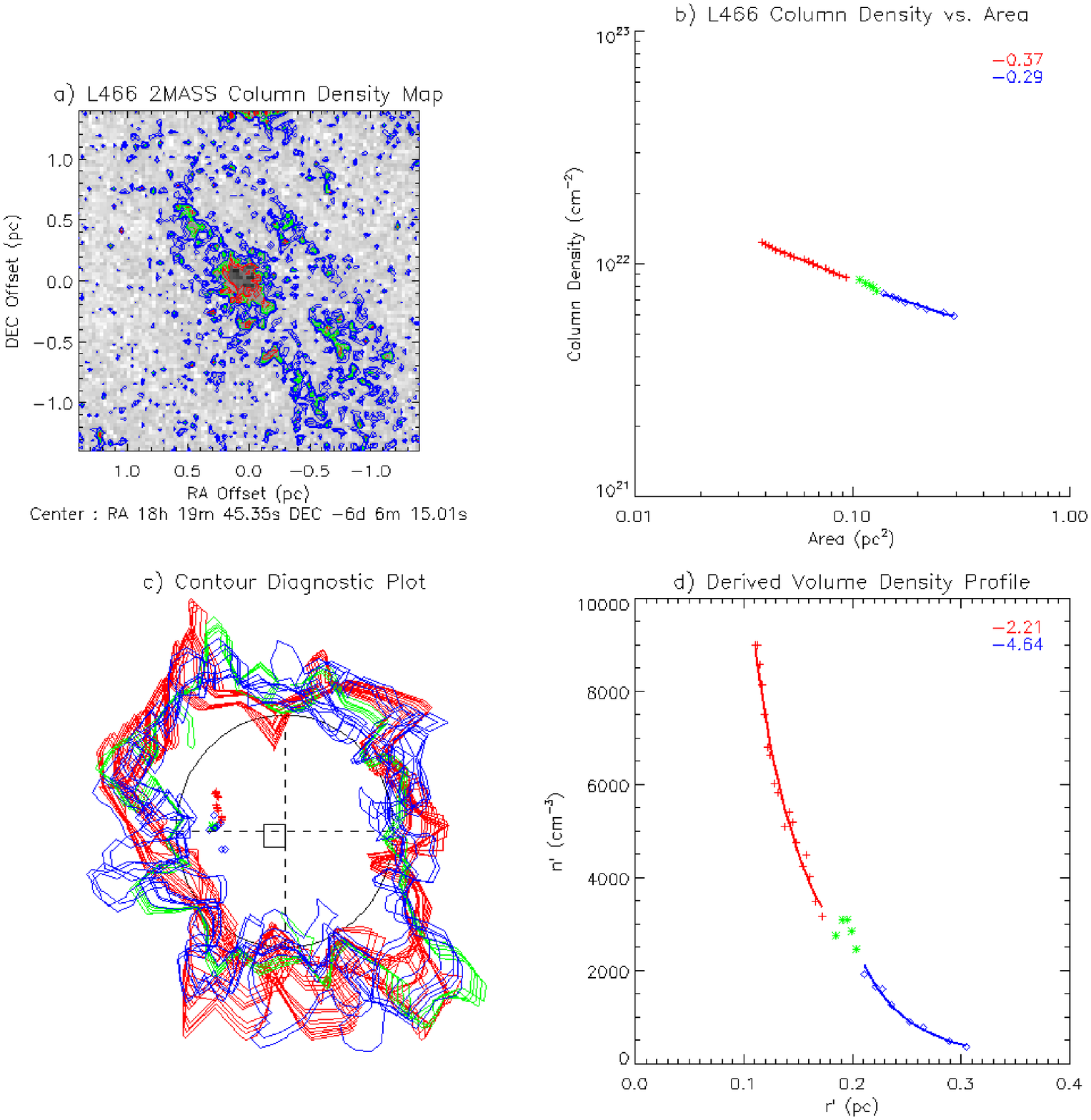}
\caption{L466 analyzed as in Figure \ref{cfig5}
\label{cfig8}}
\end{figure}

The cloud cores examined so far have all belonged to the Ophiuchus and Perseus complexes and have been surrounded by neighboring clumps which prevented us from measuring the density profiles in the outermost regions of the clouds. Furthermore, they have all exhibited a very similar power-law behavior, while originating from the same data source, which raises concerns that perhaps the way they were gridded, or the reduction method, may somehow be influencing the results. Hence, we located two cloud cores (L466, and B133) which are isolated, and employed an independent data reduction, gridding the maps to 1 arcminute beams. 

Figure \ref{cfig8} represents L466. In this case, there are no adjoining clumps, or background extinction. Thus we were able to utilize a much wider range of contours in the exterior regions of the cloud. Here, both the red and blue regions are trustworthy, while the green region corresponds to an $\epsilon$ discontinuity and is not trustworthy. Here, the red region exhibits the same behavior as the previous clouds with a power-law slope of $-2.21$. The outermost, diffuse region of the cloud also follows a power-law, but with a slope of $-4.64$. 

\begin{figure}
\includegraphics[scale=0.65]{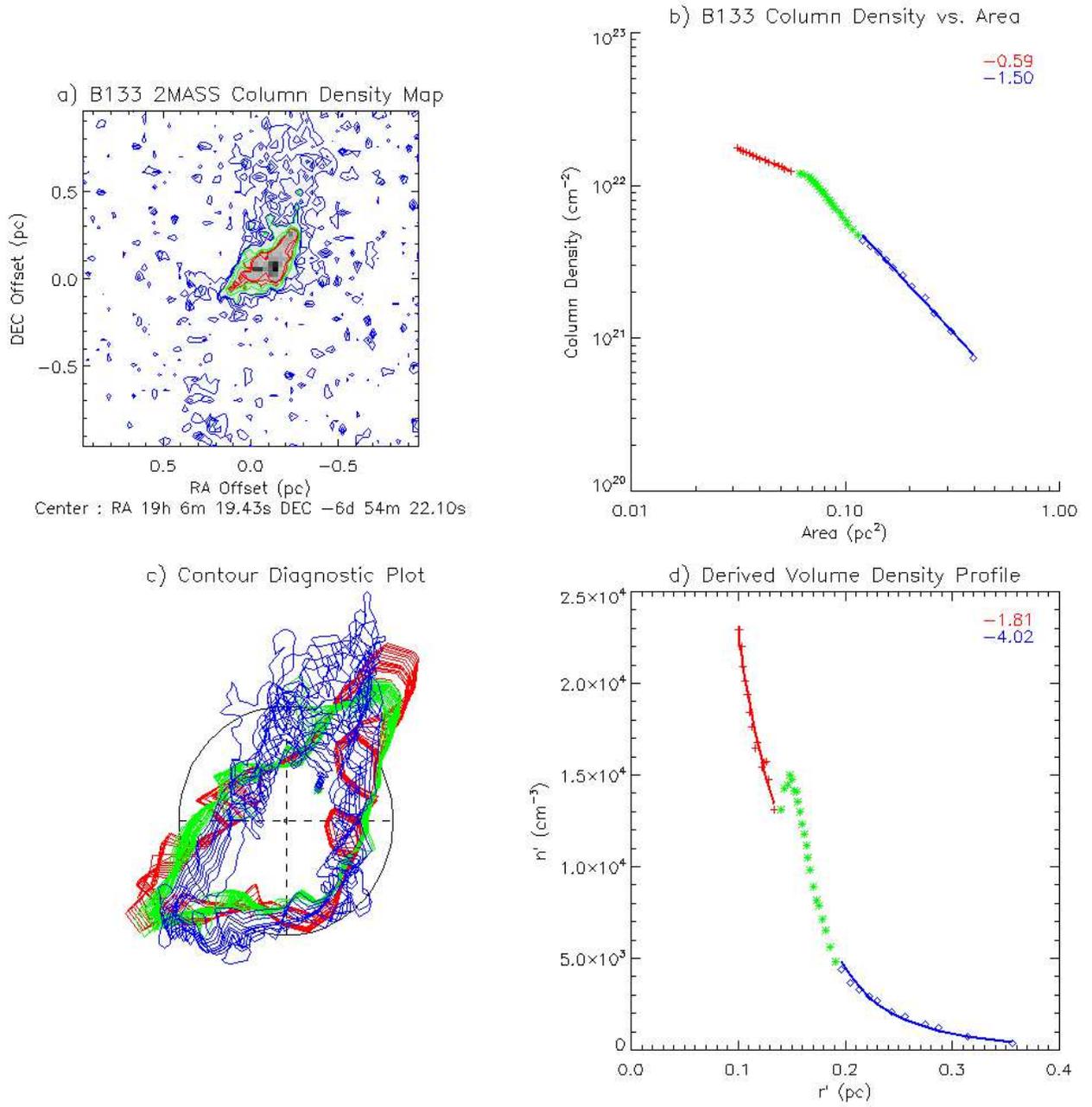}
\caption{B133 analyzed as in Figure \ref{cfig5}
\label{cfig9}}
\end{figure}

Figure \ref{cfig9} reveals that in B133 we can measure the density profile in the outermost region of the cloud. Similarly to the case of L466, there seem to be two power laws present in the red and blue regions with slopes of $-1.81$, and $-4.02$. The green region is again the site of an $\epsilon$ discontinuity (significantly larger than in L466) and cannot be trusted.

What is the meaning of the two power-laws in the two clouds (L466 and B133) where we have been able to confidently measure the radial profile in the diffuse region of the cloud? How can two distinct power laws exist within the same cloud? This kind of discontinuity can be troubling. Previous researchers have noted that attenuated power-laws can well describe such clouds, while exhibiting different localized power laws in individual regions \citep{Pineda2010}. 

It may be possible that an attenuated power law, such as that used in Equation \ref{cngnp} may accurately represent these clouds. 
There is insufficient data to fit $n_p$ to the derived profile from L466 and B133 since the innermost region is still unmeasured and $n_p$ has three free parameters ($n_0$, $\gamma$, and $a_c$). 
Using three free parameters it is not possible to derive a constrained fit for L466 and B133. 
However, the total mass of the cloud may be used to reduce this to a two-parameter problem utilizing the relationship

\begin{equation}
M = 4 \pi \int_0^{\infty} n(r) r^2 dr= 4 \pi n_0 \int_0^{\infty} \left(1+ \frac{r^2}{a_c^2}\right)^{\frac{-\gamma}{2}} r^2 dr \lc
\end{equation}
which may be integrated using the $_2F_1$ Hyper-Geometric function to yield
\begin{equation}
\label{cmasseqn}
M=\pi^{\frac{3}{2}} n_0 a_c^3 \left( \frac{\Gamma(\gamma/2 -1)}{\Gamma(2)}\right) \lc
\end{equation}
where $\Gamma$ represents the Gamma function ($\Gamma(a+1)=a!$).

Equation \ref{cmasseqn} permits us to turn the attenuated power-law fit into a two parameter problem using a cloud mass measured from the column density map. 
$\gamma$ and $n'_0$ seemed the most appropriate free parameters to use. It can be shown using the derivation in Section \ref{cderivation} that it is appropriate to use $n'_0$ and $a'_c$ along with the cloud mass even though geometric information is entangled in those parameters.

\begin{figure}
\includegraphics[scale=0.65]{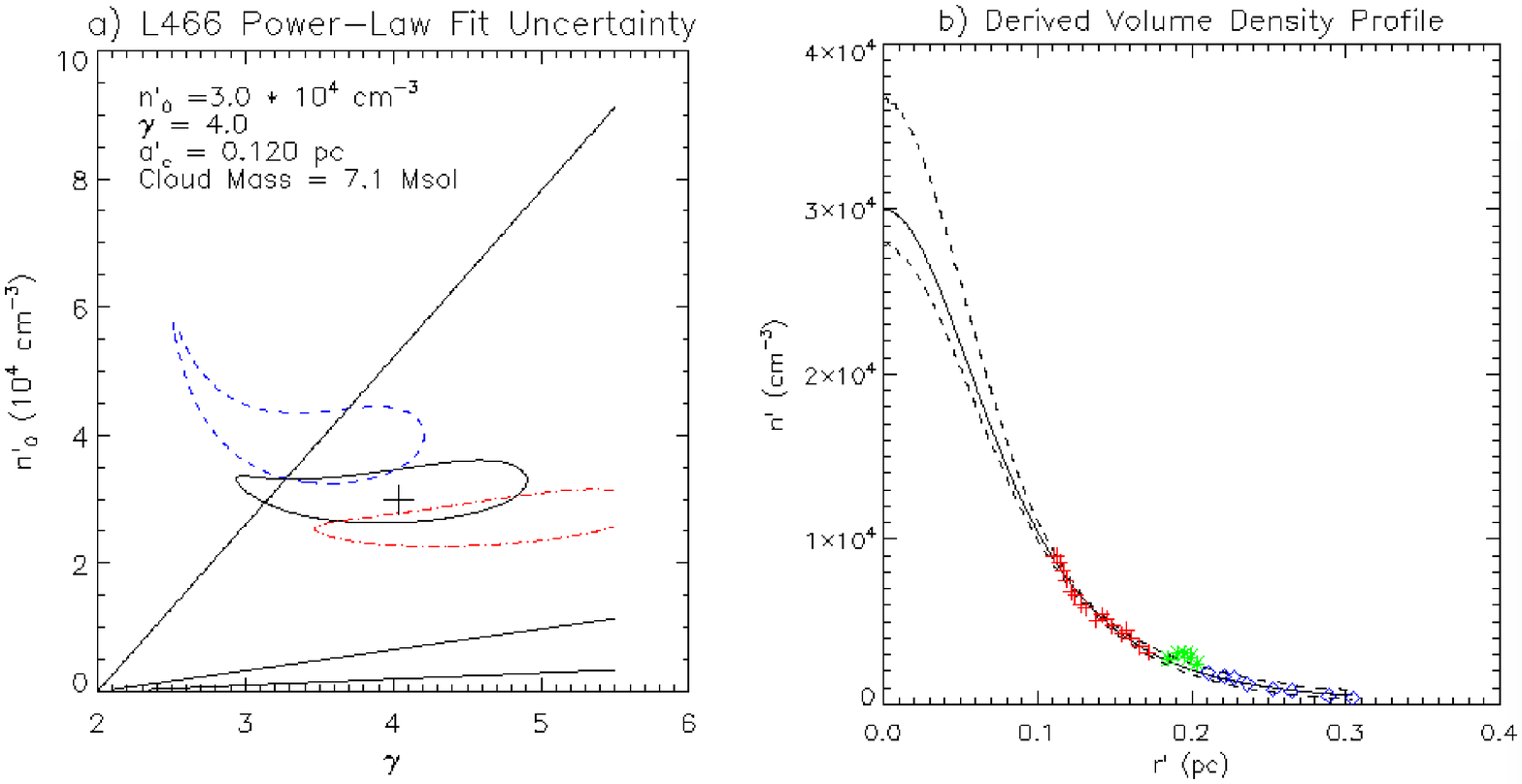}
\caption{The results of the attenuated power-law fit for L466. The green region is ignored in the calculations. a) Image of the fit residuals on a map as a function of $\gamma$ and $n'_0$. The cross represents the position of the best fit with the lowest residual. The black contour represents the $1\sigma$ uncertainty for the fit. The blue and red contours represent the uncertainties considering masses respectively $10\%$ lower, and higher than the measured value. The straight lines represent positions where $a'_c$ equals 0.1, 0.2, and 0.3 parsecs in a clockwise order. They represent the likely size of the cloud's core assuming the attenuated power-law is a correct fit. b) The derived density profile (points) along with the fitted attenuated power-law (solid line) and the $1\sigma$ uncertainty (dashed lines).
\label{cfig10}}
\end{figure}

\begin{figure}
\includegraphics[scale=0.65]{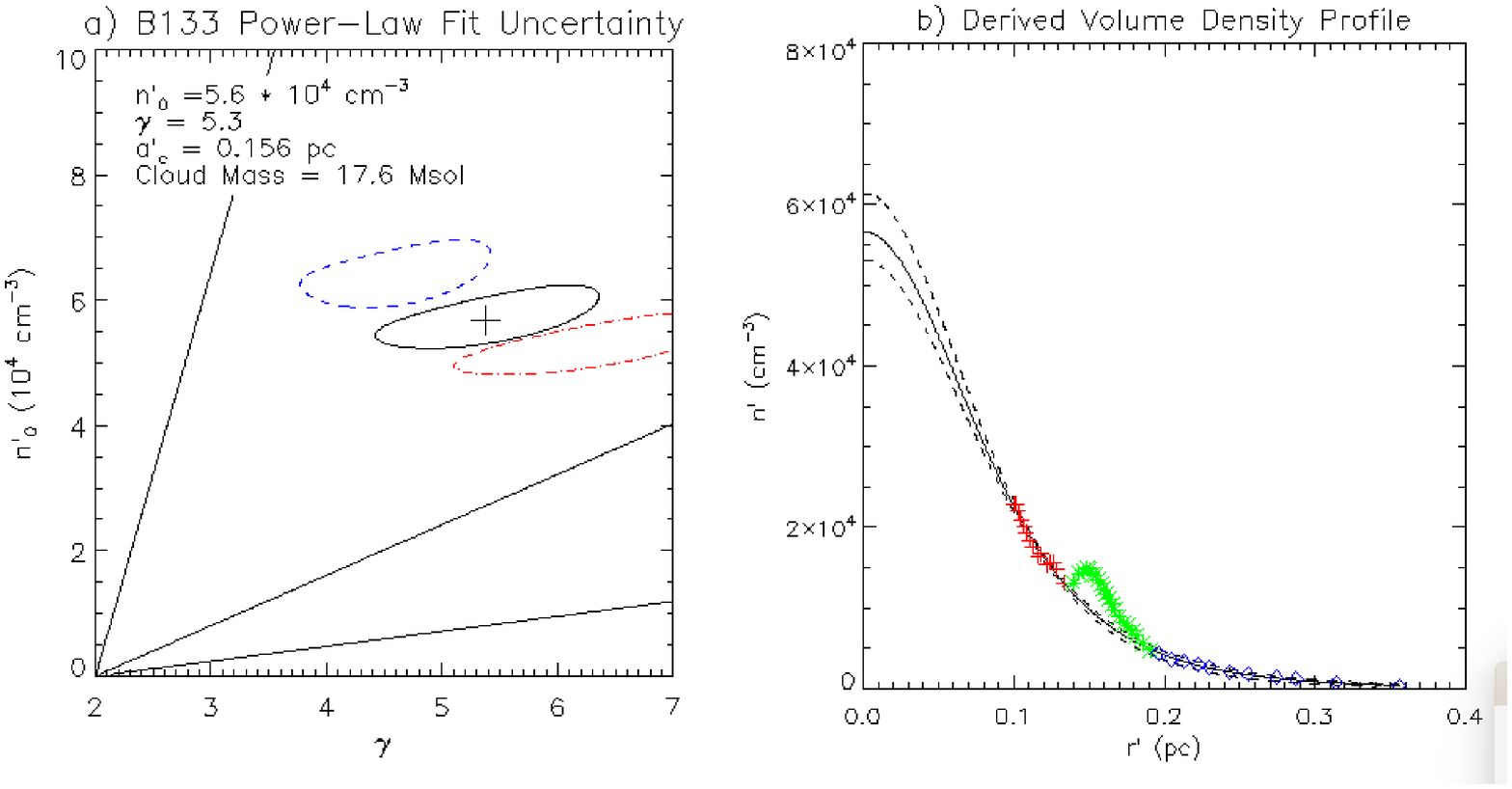}
\caption{The results of the attenuated power-law fit for B133 presented in the same form as in Figure \ref{cfig10}.
\label{cfig11}}
\end{figure}

Figures \ref{cfig10} and \ref{cfig11} present the results of such fits. The left panel in each figure represents the residuals map. The best fit for values of $n'_0$ and $\gamma$ is represented by a cross. The contours represent the $1\sigma$ uncertainties calculated from the residuals and are not necessarily symmetric. There may be some uncertainty in determining the masses of these clouds from the column density maps, due to background extinction, biases, uncertainty in the dust to gas ratio, distance estimates, and ambiguity in defining the edges of each cloud. These uncertainties may combine for several tens of percent in some cases. The blue and red contours correspond to alternate solutions with masses $10\%$ lower, and higher from the measured value to illustrate the effect. Underestimating a cloud's mass will result in a lower modelled value of $\gamma$. While mass constrains the value of $a'_c$ for any given $n'_0$ and $\gamma$, lines where $a'_c$ equals 0.1, 0.2, and 0.3 parsecs are drawn with the 0.1 parsec line closest to the vertical axis. The best-fit curves (solid line) are plotted in the right panels over the derived profiles, with the $1\sigma$ uncertainty bounds marked by dashed lines. The uncertainty bounds are not symmetric as they are not necessarily gaussian and are calculated using all possible solutions, not just the best fit.  

In both clouds, the exterior (blue) and interior (red) regions can be very well fit by an attenuated power law. It is a peculiarity of the attenuated power-law model that there may be a great deal of uncertainty in the values of $\gamma$ which, in combination with $a'_c$, and $n'_0$ can produce very similar results. The greatest uncertainty in the fit occurs in the innermost core of each cloud, as expected. There is nothing in the data which would suggest that the innermost regions of the clouds follow the fitted profile, as there is no data there. However, these fits do show that an attenuated power-law could explain why two different regions within the same cloud could appear to follow very different localized power-laws. 

\section{Previous Methods and Practices}
\label{ccomparison}

This technique may be applied to many fields of study. 
In the previous section we examined the total proton volume density radial profiles in molecular cloud cores. 
To our knowledge, no one has previously applied a geometry--independent method for measuring such volume density radial profiles.
The imposition of geometric assumptions has been especially problematic 
since molecular clouds rarely resemble simple geometries.
In fact, most studies involving the internal structure of molecular clouds have limited themselves to studying the distribution of the observed column densities.

Due to the desire to have more direct information on the internal structure of these clouds, various methods have been used to get volume density estimates despite their inherent limitations. 
The simplest and most common method involves estimating the object's shape from a column density map to arrive at an educated guess for its depth along one or more lines of sight \citep{Liu2012,Wu2012}.
This method is usually used only along a single line of sight, typically the center, due to the uncertainties in estimating an object's three-dimensional shape.
This method yields only the mean volume density along the line of sight, and is directly influenced by geometric assumptions. 
It yields no information on how volume density varies within a cloud.

Understanding the internal structure of clouds allows us to determine the relevant physics and chemistry. 
Many studies have sought to do this by comparing an assumed cloud geometry to an observed column density map in order to produce a best-fit radial profile function.
Spheres and ellipsoids are most commonly assumed.
Early examples are the Bonnor-Ebert Sphere \citep{Bonnor1956, Ebert1955} or the study by \cite{Arquilla1985}. More recent examples of this methodology are found in \cite{Pineda2010}, and \cite{Dapp09}. 
The primary advantage of this method is that it yields a volume density radial profile. 
However, it is often difficult to reconcile the idealized geometric assumptions with the actual objects studied. 
\cite{Alves2001a} found that  B68 seemed to provide an excellent fit to the Bonnor-Ebert Sphere in their survey. 
Any deviations from the assumed geometry manifest themselves in the derived radial profiles as bias in ways that are often unpredictable, and are thus frequently ignored or simply misinterpreted as uncertainties. 
We avoid such biases by discarding assumptions on the object's shape.
As a result, the afore-mentioned $\epsilon$ discontinuities become readily apparent and allow us to avoid regions where the self-similarity assumption fails.
		
\section{Conclusions}
\label{cdiscussion}
	This paper has presented a novel new method for determining the forms of the radial volume density profiles of objects such as molecular cloud cores without making assumptions about their geometry. While the method has been applied here only to dust extinction maps of molecular clouds, it is highly generalized and may be applied to any objects, and any observable quantities that satisfy the assumptions in Section \ref{cassumptions}. Those assumptions may be briefly summarized as requiring that the object can be described using a single radial profile function as well as the validity of Equation \ref{csimplen}. As such, this method may be widely useful in a number of fields. The method relies on using only a column density map, which necessarily cannot uniquely define a three-dimensional object whose geometry is unknown. It is a fortunate mathematical peculiarity that makes this method possible in that all of the object's geometric information, and no information about the form of the radial profile function are embedded into two dependent scalars ($G$ and $\chi$). These constants scale the derived $n'(r')$ profile to the cloud's original function $n(r_{rc})$. Values for $G$ and $\chi$ can only be determined with knowledge of an object's geometry.  However, the form of the derived radial profile can be derived accurately independent of geometry within the bounds of the assumptions in Section \ref{cassumptions}.
	
	Those methods which rely on geometric assumptions necessarily introduce an often significant, yet difficult to predict bias due to deviations from an idealized geometry. Our method yields the maximum amount of information attainable without the introduction of such a bias. 
Our method is limited in several ways, the chief of which is the presence of $\epsilon$ discontinuities which arise due to variations in contour shapes as a result of the failure of assumption 2. Figure \ref{cfig4}b 
a cloud core for which self-similarity is satisfied only its the outer-most and central regions.
Regions of the derived profile immediately interior to the $\epsilon$ discontinuities are affected, while exterior regions and those sufficiently far to the interior of the discontinuities still maintain their form. 
If one were to assume the simulated object in Figure \ref{cfig4}a were a sphere, the $\epsilon$ discontinuities would still be presented, but manifested in a less obvious, and more unpredictable manner. Despite its limitations, our method presents the best option for discerning the form of the radial profile in those situations where it is suitable for use.
	
	Section \ref{cderivation} presents an analytic derivation from basic principles, while Section \ref{cnumeric} bolsters the derivation through tests using simulated data. All the clouds studied here exhibited similar behavior with power laws present in each. All clouds exhibited power-laws with similar slopes ranging from $-1.8$ to $-2.5$ in their middle regions. We believe that these power-laws are not artifacts of our method since analytic derivation, and numeric simulations show that the technique should be capable of properly deriving any kind of radial profile function. 

	We chose to demonstrate the technique using 2MASS data on molecular cloud cores.
The total proton density is of particular interest as it contains information on support mechanisms as well the formation rates of molecules such as H$_2$, or $^{13}$CO. 
While we did not display the data for all clouds studied in this paper, it seems that quite often there are regions where the cloud's column density contours exhibit remarkable self-similarity accompanied by sharp changes in contour shape ($\epsilon$ discontinuities) between regions. 
		
	Our method shows that there does not appear to be a gradual change in the derived local power laws, but rather sudden shifts where the interior profile may follow a power law of $~2$, followed by an $\epsilon$ discontinuity, and a much steeper power law of $~4$ in the exterior regions. That this sudden break accompanies a stark shift in contour shapes is intriguing. With only two isolated clouds, there is insufficient data with which to draw general conclusions, and that is beyond the scope of demonstration of the technique. 
If it can be proven that this is a common characteristic of isolated molecular cloud cores and is not some kind of artifact of the column density maps or the technique for deriving radial profiles, then is there a real effect which produces a sudden change in the density behavior of these clouds. One possibility may be that the properties of the dust particles change at lower densities, thus producing a sharper drop in observed extinction, or one of the cloud's support mechanisms ceases to be effective at a certain point leading to a steeper drop in density in the exterior. It may also be possible that the interior of the cloud is undergoing gravitational collapse while the exterior is not. 
We can say at the present time that we have no reason to believe that this change in local power laws is due to biases within the data or our reduction technique within the bounds of the limitations discussed throughout the paper as we have tested the technique using a variety of geometries, profile forms, beam widths, and reduction techniques. There is nothing in the analytical derivation suggesting that such a phenomenon should be produced as a side-effect. Furthermore, while previous studies have found that the localized power law seems to change within different regions of individual cloud cores, these studies may not have been able to discern how sharp the change is due to their use of geometric assumptions which obscure the manifestation of $\epsilon$ discontinuities.

\section{Acknowledgements}

This publication makes use of data products from the Two Micron All Sky Survey, which is a joint project of the University of Massachusetts and the Infrared Processing and Analysis Center/California Institute of Technology, funded by the National Aeronautics and Space Administration and the National Science Foundation.
We wish to thank Jorge Pineda for useful discussions.
This research was carried out in part at the Jet Propulsion Laboratory operated for NASA by the California Institute of Technology.
We thank the Hayden Planetarium, and Rebecca Oppenheimer in particular for generously providing a conducive environment in which a portion of this research was carried out. We appreciate the very insightful comments and suggestions from the anonymous reviewer that significantly improved this paper.

\end{document}